\documentclass[reprint,amsmath,amssymb,prl,aps,superscriptaddress,longbibliography]{revtex4-1} 

\usepackage{color}
\usepackage{graphicx}
\usepackage{bm}
\usepackage{amsmath,amssymb}
\usepackage{physics}
\usepackage{mathtools}
\usepackage{multirow}
\usepackage{bbm}
\usepackage{xspace}
\usepackage{braket}
\usepackage{bbold}
\usepackage{xcolor}
\usepackage{upgreek}
\usepackage{tabularx}

\AtBeginDocument{\let\oldcontentsline\contentsline}
\newcommand{\notoccontentsline}[4]{\oldcontentsline{}{}{}{}}
\newcommand{\droptocpage}{\addtocontents{toc}{\let\protect\contentsline\protect\notoccontentsline}}
\newcommand{\incltocpage}{\addtocontents{toc}{\let\protect\contentsline\protect\oldcontentsline}}

\newcommand*\diff{\mathop{}\!\mathrm{d}}

\begin{document}

\title{Long-lived Bell states in an array of optical clock qubits}
\author{Nathan Schine}
\thanks{These authors contributed equally to this work.}
\affiliation{JILA, University of Colorado and National Institute of Standards and Technology, and Department of Physics, University of Colorado, Boulder, Colorado 80309, USA}
\author{Aaron W. Young}
\thanks{These authors contributed equally to this work.}
\affiliation{JILA, University of Colorado and National Institute of Standards and Technology, and Department of Physics, University of Colorado, Boulder, Colorado 80309, USA}
\author{William J. Eckner}
\affiliation{JILA, University of Colorado and National Institute of Standards and Technology, and Department of Physics, University of Colorado, Boulder, Colorado 80309, USA}
\author{Michael J. Martin}
\affiliation{Los Alamos National Laboratory, Los Alamos, New Mexico, USA}
\author{Adam M. Kaufman}
\affiliation{JILA, University of Colorado and National Institute of Standards and Technology, and Department of Physics, University of Colorado, Boulder, Colorado 80309, USA}
	    
\date{\today}
\begin{abstract}
The generation of long-lived entanglement on an optical clock transition is a key requirement to unlocking the promise of quantum metrology. Arrays of neutral atoms constitute a capable quantum platform for accessing such physics, where Rydberg-based interactions may generate entanglement between individually controlled and resolved atoms. To this end, we leverage the programmable state preparation afforded by optical tweezers along with the efficient strong confinement of a 3d optical lattice to prepare an ensemble of strontium atom pairs in their motional ground state. We engineer global single-qubit gates on the optical clock transition and two-qubit entangling gates via adiabatic Rydberg dressing, enabling the generation of Bell states, $\ket{\psi} = \frac{1}{\sqrt{2}}\left(\ket{gg} + i\ket{ee}\right)$, with a fidelity of $\mathcal{F}= 92.8(2.0)$\%. For use in quantum metrology, it is furthermore critical that the resulting entanglement be long lived; we find that the coherence of the Bell state has a lifetime of $\tau_{bc} = 4.2(6)$ s via parity correlations and simultaneous comparisons between entangled and unentangled ensembles. Such Bell states can be useful for enhancing metrological stability and bandwidth. Further rearrangement of hundreds of atoms into arbitrary configurations using optical tweezers will enable implementation of many-qubit gates and cluster state generation, as well as explorations of the transverse field Ising model and Hubbard models with entangled or finite-range-interacting tunnellers.

\end{abstract}
	
\maketitle

\droptocpage

\begin{figure*}
	\centering
	\includegraphics{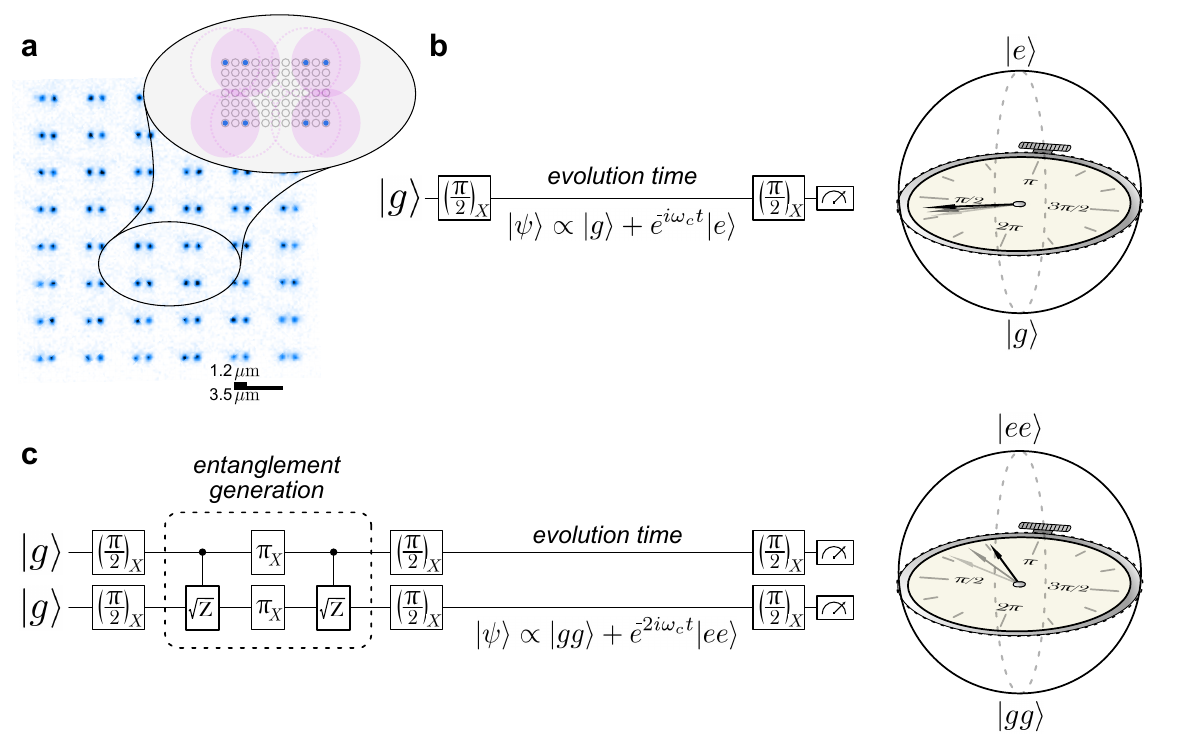}
	\caption{\textbf{Quantum-enhanced optical frequency metrology. a)} A tweezer array defines a 2d array of 1.2 $\mu$m separated doublets stochastically loaded and subsequently transferred into a 3d optical lattice (average image, blue). \textbf{Inset} Two atoms within a doublet (blue, filled circles) are two lattice sites apart (gray, open circles), placing them well within the Rydberg blockade radius of each other (pink). Doublets are separated by 6 lattice sites (3.4 $\mu$m) to preclude stray interactions between them. \textbf{b)} A standard clock measurement implements a Ramsey interferometer, in which each atom is prepared in an even superposition of the ground ($\ket{g}$) and clock ($\ket{e}$) states $\ket{\psi} = \frac{1}{\sqrt{2}}\left(\ket{g}+\ket{e}\right)$, and the clock interrogation proceeds by measuring the phase accrued by the atoms during the evolution time. On the Bloch sphere, this phase evolution corresponds to the Bloch vector precessing around the equator, acting like the hand on a stopwatch that revolves once every $\frac{2\pi}{\omega_c} = 2.3$ fs, the inverse of the energy difference between $\ket{g}$ and $\ket{e}$; comparing the ticking of this stopwatch to a laser's oscillation is at the heart of an optical clock. \textbf{c)} With multiple atoms, additional laser pulses may prepare an entangled state which is better for metrology. Here, the Bell state $\ket{\psi} = \frac{1}{\sqrt{2}}\left(\ket{gg}+i\ket{ee}\right)$ is prepared by two controlled-$\sqrt{Z}$ phase gates and global single qubit gates. The phase between $\ket{gg}$ and $\ket{ee}$ evolves twice as fast during the evolution time, yielding a stopwatch that ticks twice as fast as for a single atom. The faster ticking of states like these holds the promise of higher stability and higher bandwidth optical clocks.}
	\label{fig:intro}
\end{figure*} 

As the essential resource in quantum science, quantum entanglement enables a broad set of applications in computing, cryptography, and material science, to name a few. One powerful application arises in metrology, where greater sensitivity and higher bandwidth sensors are afforded by the properties of entangled multi-particle quantum states~\cite{kitagawa1993squeezed,wineland1994squeezed, sackett2000experimental, meyer2001experimental, esteve2008squeezing, burd2019quantum, tse2019quantum, megidish2019improved, backes2021quantum}. Combining such enhancements with state-of-the-art time and frequency metrology~\cite{ludlow_optical_2015,ushijima2015cryogenic,huntemann2016single,BACON2021frequency,mcgrew2018atomic,marti_imaging_2018,origlia_towards_2018, oelker2019demonstration,brewer2019al,sanner2019optical,BACON2021frequency,dorscher2021optical} --- namely, optical atomic clocks --- has been a defining goal in this field of quantum metrology; the construction of a quantum-enhanced optical clock has broad implications for geodesy~\cite{mehlstaubler2018atomic,takamoto2020test}, gravitational wave detection~\cite{kolkowitz2016gravitational, hu2017atom, abe2021matter}, and the search for physics beyond the standard model~\cite{safronova2018search,sanner2019optical}. 

A variety of approaches exist for creating metrologically useful entanglement. In neutral atom optical lattice clocks, a number of methods have been proposed using cavity QED, Rydberg interactions, or collisional interactions~\cite{leroux2010implementation, gil_spin_2014,strobel2014fisher, zeiher2016many,cox2016deterministic,engelsen2016engineering,lewis2018robust,huang2020self}--- indeed recently, spin squeezed states on an optical clock transition have been generated using collective cavity QED interactions~\cite{pedrozo2020entanglement}.  In trapped ions, proposals and implementations of entanglement on optically separated qubits rely on spin-spin interactions mediated by Coulombic crystal modes, allowing efficient entanglement generation and GHZ states with as many as 24 ion optical qubits~\cite{pogorelov2021compact}. Given the control and measurements possible in modern atomic clock architectures, these systems offer the possibility of merging quantum information concepts with precision measurement. Utilizing entangling gates and protocols from quantum information science broadens the set of realizable quantum states, enabling new protocols for quantum metrology, while quantum error correcting codes have opened routes for enhanced quantum sensing in the presence of noise and imperfect quantum resources~\cite{toth2014quantum, kessler2014quantum, zhou2018achieving,  kaubruegger_variational_2019, layden2019ancilla, kaubruegger2021quantum, shettell2021practical}. In the context of optical atomic clocks, a number of recent protocols have advanced this union as a route to quantum-enhanced measurements~\cite{toth2014quantum, kaubruegger_variational_2019, kaubruegger2021quantum,marciniak2021optimal}, and, in a very recent demonstration, variational optimization of phase sensitivity at optical frequencies was implemented on a 26 ion quantum processor~\cite{marciniak2021optimal}. 

This theoretical and experimental momentum emphasizes the promise of combining control with scalability, in order to produce large-scale, entangled clocks amenable to versatile control and measurement schemes~\cite{kessler2014quantum, kaubruegger_variational_2019, pezze2020heisenberg}. An outstanding challenge in this regard is to pair these capabilities with long-lived quantum coherence, so that many-particle states can be fruitfully leveraged for quantum-enhanced measurements at long interrogation times. A promising candidate for realizing this synergy is the recently demonstrated tweezer clock, which not only supplies long-lived atomic coherence and high stability~\cite{norcia_seconds-scale_2019, madjarov_atomic-array_2019, young_half_2020}, but also large atom number~\cite{young_half_2020, madjarov_atomic-array_2019, scholl2021quantum, ebadi2021quantum}, high-fidelity Rydberg interactions~\cite{levine_parallel_2019, graham_rydberg_2019, madjarov2020high}, and microscopic control to engineer precisely tailored quantum systems.

\begin{figure*}
	\centering
	\includegraphics[width=\textwidth]{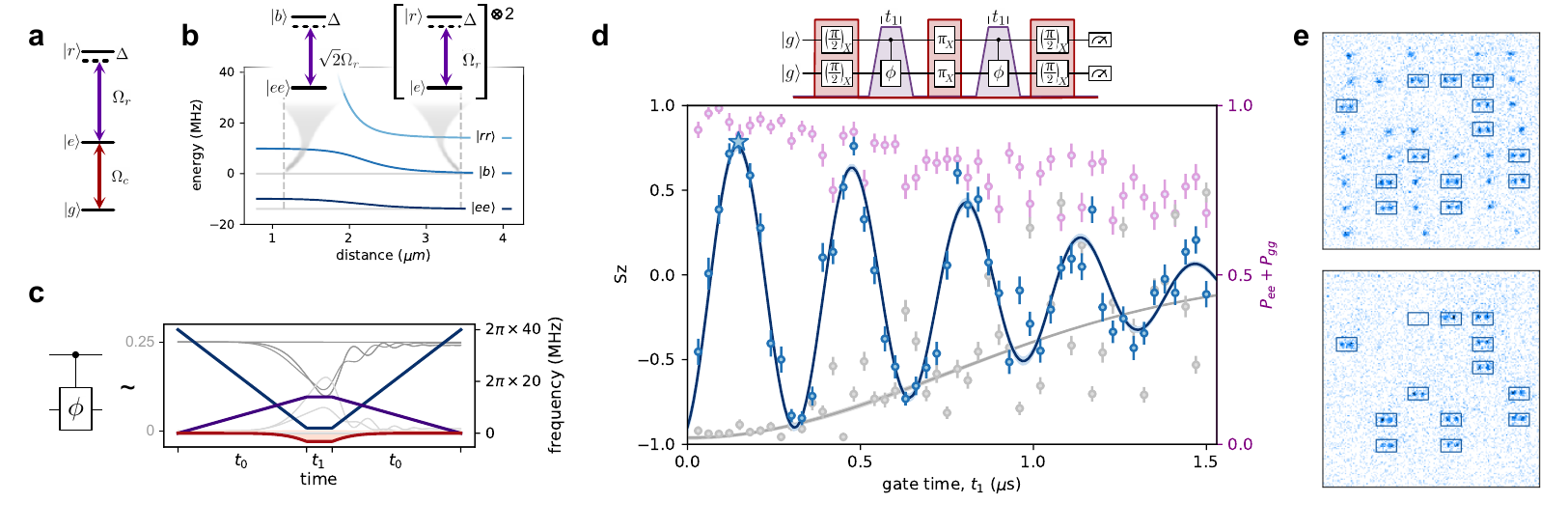}
	\caption{\textbf{Clock qubit controlled phase gate via adiabatic Rydberg pulses. a)} The ground ($\ket{g}$) and clock ($\ket{e}$) states define the clock transition qubit, which may be driven resonantly with Rabi frequency $\Omega_c$.  To engineer entanglement, we drive the clock state to a Rydberg state ($\ket{r}$) with controllable detuning $\Delta$ and Rabi frequency $\Omega_r$. \textbf{b)} We consider the effect of interactions on the dressed two-atom $\ket{e}$-$\ket{r}$ subsystem. At large distances (e.g. inter-doublet spacing, right vertical dashed line), interactions between two Rydberg atoms are negligible. As the distance between atoms decreases (e.g. intra-doublet spacing, left vertical dashed line), Rydberg blockade causes an effective reduction of the Hilbert space to $\ket{ee}$ and $\ket{b} = \frac{1}{\sqrt{2}}(\ket{er}+\ket{re})$. This modifies the lightshift on the $\ket{ee}$ state, providing the mechanism for entangling operations. \textbf{c)} By ramping both the detuning (blue) and Rabi frequency (purple), we adiabatically follow the dressed $\ket{ee}$ state to small detuning to provide a large entangling energy (red) and a fast controlled phase gate. A master equation model (see SI) tracks populations (left axis) within (dark gray) and outside (light gray) the two-qubit computational states as the gate progresses. The Rydberg state is populated only during the middle of the gate. \textbf{d)} To remove the effects of the large single atom lightshift and provide the dynamics of only controlled phase gates, we implement a spin echo gate sequence (shown at top, timing not to scale): two adiabatic Rydberg pulses separated by a clock $\pi$-pulse implement an effective $\hat{S}_z^2$ Hamiltonian, while two surrounding clock $\pi/2$-pulses convert the resulting phase shifts into population oscillations between $\ket{gg}$ and $\ket{ee}$ (plotted as $S_z\equiv\langle \hat{S}_z\rangle$, blue, left axis) and nowhere else (pink, right axis). Meanwhile, non-interacting atoms (half-filled doublets, gray, left axis, plotted as $2\langle \hat{S}_z \rangle$ to match vertical scales) undergo a $2\pi$ pulse and return to the ground state. For longer pulses, the spin echo sequence fails to cancel the large single atom lightshift, resulting in relaxation towards $S_z = 0$ and population leaving the $\ket{gg}$-$\ket{ee}$ subspace. All error bars are the s.e.m. from averaging over the array and 30 repetitions of the experiment, and fit error bands represent 1$\sigma$ uncertainty. \textbf{e)} We display an illustrative pair of images corresponding to the first peak of the interaction oscillation, shown as a star in panel \textbf{d}. The first image enables identification of empty, half-filled, and fully-filled doublets, while the second image identifies clock state atoms after the gate sequence.}
	\label{fig:gate}
\end{figure*}

In our previous work, we demonstrated state-of-the-art relative stability and atomic coherence using a tweezer array clock~\cite{norcia_seconds-scale_2019,young_half_2020}. Here, we advance our atomic control of the clock transition to elevate it into a high-fidelity qubit. We then implement entangling gates between these qubits through Rydberg excitations out of the excited qubit state, which realizes controllable Ising-type interactions~\cite{wilk2010entanglement, isenhower2010demonstration, labuhn2016tunable, levine_parallel_2019, graham_rydberg_2019,madjarov2020high}. 

With these capabilities, we achieve high-fidelity long-lived entanglement in an ensemble of Bell state pairs (Fig.~\ref{fig:intro}a)~\cite{leibfried2003experimental, kirchmair2009high, ballance_high_2016, gaebler_high_2016, shapira2018robust, levine_parallel_2019, graham_rydberg_2019}. We employ adiabatically resonant coupling of the qubit excited state to a Rydberg state to implement a controlled phase gate between two atomic qubits, up to single atom lightshifts~\cite{mitra_robust_2020}.  Setting the controlled phase of our gate to be $\pi/2$ yields a controlled-$\sqrt{Z}$ gate and enables the production of the Bell state $\ket{\psi} = \frac{1}{\sqrt{2}}(\ket{gg}+i\ket{ee})$, which has metrological relevance because the superposition components have a large energy separation. As such, clocks based on such states accrue phase more quickly than clocks based on unentangled atoms (Fig.~\ref{fig:intro}b,c)~\cite{meyer2001experimental, kessler2014heisenberg}. We determine the fidelity with which we produce Bell states to be $\mathcal{F}=92.8(2.0)\%$ over a 6$\times$8 array of atom pairs. To better understand near-term prospects for quantum metrological enhancement, we study the lifetime of these states in several ways. We observe the parity oscillation contrast and resulting fidelity to decay with a Gaussian $1/e$ time constant of $\sigma_{F} = 407(13)$ ms~\cite{sackett2000experimental}. However, the loss of atom-laser coherence can occur without the loss of atom-atom coherence, or, in this case, relative coherence between distinct Bell states. Taking inspiration from measurements of atomic coherence that reject the relative phase noise of the interrogation laser~\cite{chou_frequency_2010, marti_imaging_2018, young_half_2020}, we measure Bell state parity and single-atom $\hat{S}_z$ correlations to infer a $1/e$ exponential Bell state coherence time of $\tau_{bc} = 4.2(6)$ seconds. 

Our experiments begin with a tweezer-defined doublet array stochastically filled with $^{88}$Sr atoms (Fig.~\ref{fig:intro}a). The atoms are implanted into a single plane of a colocated 3d optical lattice, imaged, and subsequently cooled into their 3d motional ground state in order to achieve high-fidelity atomic control (see methods). We then apply a sequence of global laser pulses that drive the $^1\text{S}_0$ ($\ket{g}$) $\leftrightarrow$ $^3\text{P}_0$ ($\ket{e}$) clock transition resonantly with Rabi frequency $\Omega_c$, and the $\ket{e} \leftrightarrow \ket{r}$ Rydberg transition (to 5s40d $^3\text{D}_1$, $m_j=0$) with Rabi frequency $\Omega_r$ and positive detuning $\Delta$ (see Fig.~\ref{fig:gate}a). Finally, we image excited clock state atoms.

We implement a controlled phase gate for two clock-transition qubits using adiabatic Rydberg pulses, following the proposal of Mitra, et al.~\cite{mitra_robust_2020}. Each Rydberg pulse ramps $\Omega_r$ and $\Delta$ so that clock state atoms adiabatically follow the instantaneous dressed clock-like eigenstate in the presence of the Rydberg coupling laser. This lightshifts the energy of a single clock state atom by $E^{(1)}_{LS} = \frac{\hbar}{2}\left(-\Delta + \sqrt{\Omega_r^2+\Delta^2} \right)$. The energy shift of two non-interacting clock state atoms is $2E^{(1)}_{LS}$. As shown in Fig.~\ref{fig:gate}b at close distances, such as two atoms within a single doublet, Rydberg blockade prevents the excitation of both atoms to the Rydberg state. The Hilbert space of two clock-state atoms then reduces to an effective two-level system with states $\ket{ee}$ and $\ket{b} = \frac{1}{\sqrt{2}}(\ket{er}+\ket{re})$ driven by a collectively enhanced Rabi frequency $\sqrt{2}\Omega_r$ and detuning $\Delta$. This lightshifts the energy of the $\ket{ee}$ state by $E^{(2)}_{LS} = \frac{\hbar}{2}\left(-\Delta + \sqrt{2\Omega_r^2+\Delta^2}\right)$. The difference $\kappa = E^{(2)}_{LS} - 2E^{(1)}_{LS}$ is named the entangling energy, as it sets the energy scale of entangling operations~\cite{mitra_robust_2020}.	

This adiabatic Rydberg pulse implements a controlled phase gate on the two-qubit states of a filled doublet, once the single atom phase is removed. That is, the Rydberg pulse realizes the Hamiltonian 
\begin{equation}
H_r = E^{(1)}_{LS}\hat{S}_z  + \kappa\ket{ee}\bra{ee}
\end{equation}
where $\hat{S}_z = \mathbb{1}\otimes\hat{\sigma}_z/2 + \hat{\sigma}_z/2\otimes\mathbb{1}$, $\hat{\sigma}_z$ is the Pauli $z$ matrix, and $\mathbb{1}$ is the identity matrix acting on the $\ket{e}$, $\ket{g}$ basis. We only consider single- and two-particle effects because the spacing between doublets is chosen to be large compared to the Rydberg interaction range (see Fig.~\ref{fig:intro}a). By implementing two of these Rydberg pulses separated by a $\pi$-pulse, we remove the single atom phase from $\hat{S}_z$, so the full sequence can be interpreted as containing two controlled phase gates, $U_{c\phi} = \ket{gg}\bra{gg}+\ket{ge}\bra{ge}+\ket{eg}\bra{eg}+e^{i\phi}\ket{ee}\bra{ee}$ with phase $\phi = - \int \kappa \diff t$ (see Fig.~\ref{fig:gate}c). Together, these three pulses implement Ising-type interactions $H_{\text{eff}} = \kappa \hat{S}_z^2$.  We note that while fixed, large-detuning Rydberg dressing would produce equivalent coherent dynamics, adiabatically following the instantaneous eigenstates to resonance increases the ratio of interactions to dissipation, improving the gate fidelity (see SI). 

We reveal the dynamics of the controlled phase gates by including them inside a Ramsey interferometry sequence (see Fig.~\ref{fig:gate}d, top), wherein the gate appears between two $\pi/2$-pulses. The first pulse establishes a superposition of well-defined phase between $\ket{g}$ and $\ket{e}$ for each atom. The spin-echoed Rydberg pulses introduce an additional phase $\phi$ for the $\ket{ee}$ and $\ket{gg}$ components, and the final $\pi/2$-pulse closes the interferometer, converting those phases into an oscillation between $\ket{gg}$ and $\ket{ee}$ and removing population from $\ket{ge}$ and $\ket{eg}$. The two $\pi/2$-pulses convert the effective Ising interactions of the middle three pulses into an $S_x^2$ interaction which rotates $\ket{gg}$ into $\ket{ee}$. As shown in Fig.~\ref{fig:gate}d, for short gate times, we observe these dynamics via $S_z \equiv \langle \hat{S}_z \rangle$ and $P_{gg}+P_{ee}$ population measurements in filled doublets. For longer gate times, we observe that the failure of the spin echo causes decay of $S_z$ oscillations as revealed in the dynamics of single atom $S_z$ relaxing to zero. Furthermore, this causes population to leave the $\ket{gg}$-$\ket{ee}$ subspace. 

\begin{figure}
	\centering
	\includegraphics[width=\columnwidth]{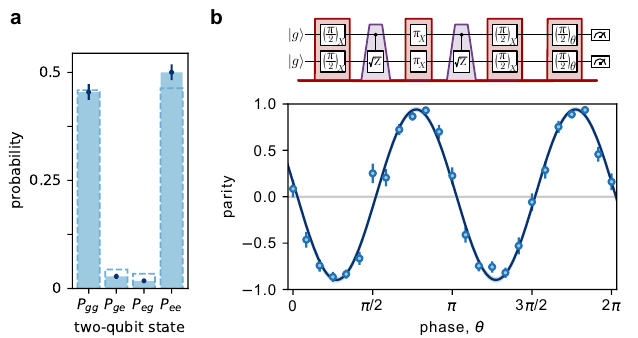}
	\caption{\textbf{Bell state fidelity.}  \textbf{a)} The array-averaged population fractions of two-qubit states reveal $P_{gg}+P_{ee} = 0.955(28)$ (blue) after correcting for independently calibrated SPAM errors (3.5\% total loss and infidelity). Uncorrected population measurements are shown by dashed outlines. Ideal gate operation would yield $P_{gg}=P_{ee}=0.5$. Error bars are s.e.m. from averaging over the array and 100 repetitions of the experiment. \textbf{b)} We measure the coherence, $C$, between $\ket{gg}$ and $\ket{ee}$ via the parity oscillation contrast. An analysis $\pi/2$-pulse about a variable axis is implemented by the composition of a single variable $\hat{\sigma}_z$ gate and a static $\pi/2$-pulse and produces two-particle states that oscillate in parity $\hat{\Pi} = \hat{\sigma}_z^i\hat{\sigma}_z^j$. The resulting data (blue points, SPAM corrected, error bars are s.e.m from averaging over array and 15 repetitions of the experiment) are fit by a sinusoid (blue curve, fit error band is represents 1$\sigma$ uncertainty). The amplitude of the sinusoidal fit yields $C = 0.919(28)$. The uncertainty arises in approximately equal parts from fit parameter estimation uncertainty and systematic imaging loss fluctuations.}
	\label{fig:parity}
\end{figure}

\begin{figure*}
	\centering
	\includegraphics[width=\textwidth]{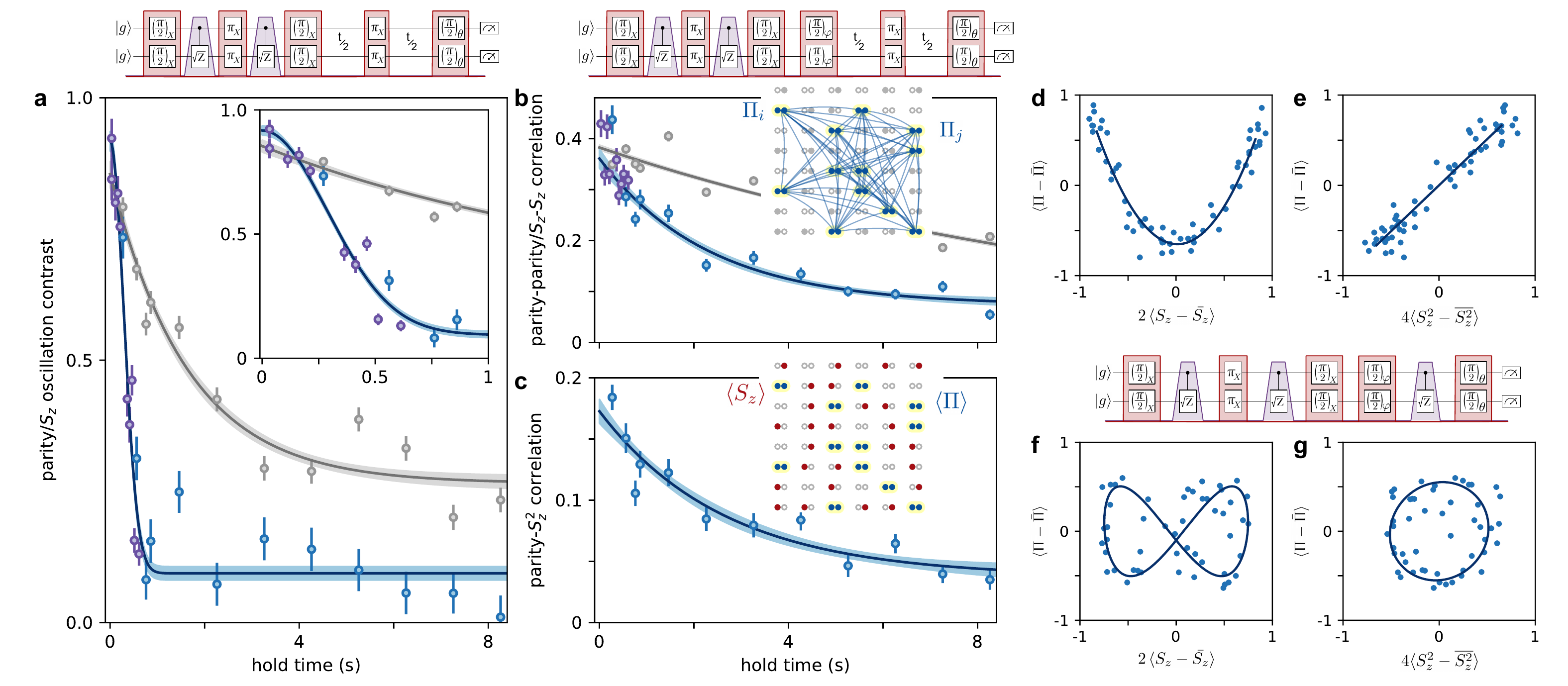}
	\caption{\textbf{Bell state coherence time.} \textbf{a)} We estimate the Bell states' coherence time through the decay of the parity oscillation contrast after a variable hold time (purple, first experimental schematic). We repeat this experiment while simultaneously preparing single atoms in a superposition state (via an additional $\pi/2$-pulse about an axis at $\varphi = 3\pi/4$, second experimental schematic). The resulting array-averaged single atom spin oscillation contrast decays exponentially with a time constant $\tau_{S_z}=1.6(2)$ s (dark gray), while the Bell states' parity oscillation contrast decays with a Gaussian with a 1/$e$ time constant $\sigma_{pc} = 407(13)$ ms (blue). For these data, a clock $\pi$-pulse is employed half way through the hold time (see methods). The inset shows detail at short hold times and shares axes with the main panel. Error bars represent the 1$\sigma$ confidence interval of the fitted sinusoidal amplitude. \textbf{b)} We access the Bell state coherence time without dependence on atom-laser coherence via parity-parity correlations (see main text). This decays with an exponential time constant of $\tau_{\Pi\Pi} = 2.3(4)$ s, from which we infer a Bell state coherence time of $\tau_{bc}^{\Pi\Pi} = 4.6(7)$ s. As an experimentally determined expectation for this, we also measure the decay rate of the $S_z$-$S_z$ correlation of single atoms (gray), which we find to be $\tau_{S_zS_z} = 12.2(5)$ s. \textbf{c)} An alternative handle on the Bell state coherence time comes from the comparison of the Bell state parity with the square of the single atom array-averaged spin projection, $S_z$. We find that this decays with an exponential time constant of $\tau_{\Pi S_z^2} = 2.7(6)$ s, from which we infer a Bell state coherence time of $\tau_{bc}^{\Pi S_z^2} = 3.4(1.1)$ s. Error bars in \textbf{b} and \textbf{c} represent the s.e.m. from averaging the array-averaged correlation over repetitions and analysis phases. \textbf{d,e)} Parametric plots of the Bell state ensemble parity against the single atom ensemble $S_z$ (\textbf{d}) and $S_z^2$ (\textbf{e}) at hold times less than a second form 2:1 and 1:1 Lissajous figures with zero relative phase, i.e. a parabola and a line. \textbf{f,g)} We introduce a relative phase difference between the entangled and unentangled ensembles with a third controlled-$\sqrt{Z}$ pulse. This opens the 2:1 and 1:1 Lissajous figures into a figure-eight and ellipse with a relative phase of 0.94(7)$\frac{\pi}{2}$ radians. Panels \textbf{d}-\textbf{g} visually demonstrate the widely-touted frequency doubling of Bell states' precession rate compared to single atoms, while squaring the single atom signal matches the oscillation frequency. Note that the blue and gray data in panels \textbf{a}-\textbf{e} are derived from the same dataset, with 150 runs of the experiment per hold time. Fit error bands represent 1$\sigma$ uncertainty.}
	\label{fig:lifetime}
\end{figure*}

A zero-crossing of $S_z$ corresponds to the application of controlled phase gates of $\phi = \pi/2$ in under a microsecond and, after the full pulse sequence of figure~\ref{fig:gate}d, the production of the Bell state $\ket{\psi} = \frac{1}{\sqrt{2}}(\ket{gg}+i\ket{ee})$. We benchmark our gates through the fidelity, $\mathcal{F}$, with which we create these Bell states, $\mathcal{F} = \frac{1}{2}\left(P_{gg}+P_{ee}+C\right)$~\cite{sackett2000experimental}. This depends on two factors: the sum of populations in the relevant states $P_{gg}+P_{ee}$ as well as the coherence between them as accessed by the contrast $C$ in a parity oscillation measurement~\cite{sackett2000experimental}. At the first $S_z$ zero-crossing, we find $P_{gg}+P_{ee} = 0.955(28)$ or 0.922(27) without state preparation and measurement (SPAM) correction (see Fig.~\ref{fig:parity}a and SI). To extract the Bell state coherence, we add a final ``analysis'' $\pi/2$-pulse about a variable axis after completing the spin echo sequence. In figure~\ref{fig:parity}b we plot the measured parity $\Pi \equiv \langle \hat{\Pi} \rangle = \langle\hat{\sigma}_z^l\hat{\sigma}_z^r\rangle$ where the superscript denotes the left and right sites within a doublet and the brackets imply averaging over the array and repetitions of the experiment. The amplitude of a sinusoidal fit versus the analysis phase yields the parity oscillation contrast $C = 0.919(28)$ and a resulting Bell state fidelity of $\mathcal{F} = 92.8(2.0)$\% (or, without SPAM correction, $C=0.838(19)$ and $\mathcal{F}=87.1(1.6)\%$). The reported fidelity is reduced by 0.9(4)\% to account for a systematic overestimation of $P_{gg}$ due to Rydberg state loss (see SI), and the uncertainty includes population and contrast uncertainty as well as uncertainty from SPAM correction, primarily from fluctuations of the imaging loss at the percent level. 

The adiabatic Rydberg ramp speed and detuning endpoints are optimized to satisfy several competing constraints. Operating faster reduces loss from the finite Rydberg state lifetime ($10.9(4)$ $\mu$s, see SI), while operating slower results in greater adiabaticity wherein atoms more completely return to the qubit states (see SI). We balance these competing effects with ramp durations of 250-350 ns (see methods). Master equation modeling (see Fig.~\ref{fig:gate}c and SI) guided the experimental optimization of these parameters (see methods).

For metrological use, a long coherence time for the entangled state is desired. In figure~\ref{fig:lifetime}, we investigate this timescale in several ways. First, we can simply wait before measuring populations and parity oscillation contrast. While the populations decay slowly (12.3(1.0) s exponential time constant for $P_{ee}$, see SI), we find that the parity oscillation contrast decays rapidly, with a Gaussian 1/$e$ time constant of $\sigma_{pc} = 407(13)$ ms (see Fig.~\ref{fig:lifetime}a). This loss of Bell-state-laser coherence could arise due to laser noise (in the rotating frame of the atoms), rather than atomic dephasing or decoherence effects. To investigate this possibility, we prepare the ensemble of single atoms in the array into a superposition state $\frac{1}{\sqrt{2}}(\ket{g}-i\ket{e})$ while preserving the Bell states' parity via an additional $\pi/2$-pulse about an axis $\varphi=3\pi/4$ before the hold time. We find that the contrast of the array-averaged spin-projection of single atoms, $S_z$, decays with a longer exponential lifetime $\tau_{S_z}=1.6(2)$ s. These measurements include a spin-echo clock $\pi$-pulse half way through the hold time to reject static and slow dephasing processes which otherwise obscure entanglement lifetime data in repetition- and array-averaged data (see methods). Separate measurements of the single-atom Ramsey contrast decay without a spin echo find an atom-laser coherence time consistent with $\tau_{S_z}$.

At times beyond the coherence time with the laser, we investigate, in two ways, whether the Bell states retain coherence with respect to the rest of the array. First, we utilize parity-parity correlations,  $\langle(\hat{\Pi}_i-\bar{\Pi})(\hat{\Pi}_j-\bar{\Pi}) \rangle$, where $\bar{\Pi} = \langle \hat{\Pi}_i \rangle$ is the parity averaged over all sites, repetitions, and analysis phases at a given hold time, and $\langle...\rangle$ indicates an average over all choices $i$,$j$ of Bell states in a single shot of the experiment, then followed by an average over all repetitions and analysis phases. This quantity should decay at twice the rate of decay of the Bell state coherence, specifically the off-diagonal coherence of the two particle density matrix defined in the frame of the other Bell states, and it is insensitive to the loss of atom-laser coherence (see SI). We find that the parity-parity correlation decays with an exponential time constant of $\tau_{\Pi\Pi} = 2.3(4)$ s indicating a Bell state coherence time of $\tau_{bc}^{\Pi\Pi} = 4.6(7)$ s. 

The parity-parity correlation decay effectively uses the relative phase within one Bell state as a reference against which to measure the coherence of another Bell state; averaging over the array, this can be construed as comparing the evolving Bell state phase with respect to the mean phase of all other Bell states. But, rather than compare Bell states to each other, we can also compare all of them to the ensemble of non-entangled atoms. More precisely, we compare the single-experiment average parity, $\Pi$, of the Bell states to the single-experiment average spin projection, $S_z$ of single atoms. However, because the Bell state parity oscillates twice as fast with varying analysis phase as the single atom $S_z$, we must first square the single atom $S_z$ to provide a reference signal which oscillates at the same frequency as the Bell state parity (see Fig.~\ref{fig:lifetime}c-g). We construct the parity-$S_z^2$ correlation, $4\langle (\Pi - \bar{\Pi})(S_z^2-\overline{S_z^2}) \rangle$, where $\langle...\rangle$ here indicates an average over all repetitions and analysis phases, and $\bar{S_z^2} = \langle S_z^2 \rangle$, and we find that this correlation decays with an exponential time constant of $\tau_{\Pi S_z^2} = 2.7(6)$ s. This decay is sensitive to both the Bell state coherence decay as well as the single-atom $S_z$-$S_z$ correlation decay (see Fig.~\ref{fig:lifetime}b, gray, plotting $4\langle (\hat{S}_z^i - \bar{S}_z)(\hat{S}_z^j - \bar{S}_z) \rangle$), which has lifetime $\tau_{S_zS_z} = 12.2(5)$ s; from these data, we infer a Bell state coherence lifetime of $\tau_{bc}^{\Pi S_z^2} = 3.4(1.1)$ s (see SI). Importantly, these correlations do not by themselves certify the presence of a Bell state, since there are non-entangled states that could result in similar signals. However, figure~\ref{fig:parity} provides tomographic evidence for the initial generation of a Bell state, and these correlations then provide the Bell state coherence time with assumptions that certain exotic decay processes do not occur (see SI). 

The comparison of the ensemble of Bell states to the ensemble of single atoms in figure~\ref{fig:lifetime}c is a simultaneous comparison between an entangled and unentangled pair of optical atomic clocks. The parametric plots of the entangled ensemble parity against the unentangled ensemble $S_z$ and $S_z^2$ at hold times less than a second form 2:1 and 1:1 Lissajous figures with zero relative phase, i.e. a parabola and a line (Fig.~\ref{fig:lifetime}d,e). The operation of a quantum-enhanced optical atomic clock will require the estimation of the relative phase difference between reference and target clocks. To examine this, we inject a relative phase difference of $\pi/2$ with a third controlled-$\sqrt{Z}$ pulse. This opens the 2:1 and 1:1 Lissajous figures into a figure-eight and ellipse, respectively, with a relative phase of 0.94(7)$\frac{\pi}{2}$ radians measured via ellipse fitting (Fig.~\ref{fig:lifetime}f,g). Future operation of this system in a quantum-enhanced simultaneous clock comparison may require establishing an optimal initial phase difference and subsequently estimating the opening angle of the figure-eight or ellipse at an optimal interrogation time, while in-situ comparison of an entangled and unentangled ensemble opens new routes for characterizing clock systematics and retaining enhanced metrological sensitivity in the presence of laser noise~\cite{kessler2014quantum, pezze2020heisenberg}.  

The two measurements of the Bell state coherence time from correlations are mutually consistent and may be combined to yield $\tau_{bc} = 4.2(6)$ s. This, combined with our understanding of single-atom and laser decoherence mechanisms, provides an expected exponential decay time for the parity oscillation contrast of 0.69(10) s, longer than we observe in figure~\ref{fig:lifetime}a. Nevertheless, the Bell state coherence time from correlations may be compared to an experimentally determined expectation given by half the single-atom atomic coherence time, i.e. the decay rate of the $S_z$-$S_z$ correlation of single atoms, $\tau_{S_zS_z} = 12.2(5)$ s. This value is understood in terms of single-atom population loss, decay, and decoherence processes, up to an unexplained single atom lossless decoherence process with a four-minute timescale. This value is also consistent with observed Bell state population dynamics, indicating the absence of correlated loss or decay processes (see SI). The nearly factor of three reduction in the Bell state coherence time from in-situ correlation measurements could indicate the existence of an unidentified additional Bell state decoherence process with an approximately six second timescale. However, the late time behavior of the correlations are consistent with single atom measurements, and we hypothesize that the observed discrepancy is the result of technical noise that results in an additional, faster decoherence timescale at early times (see SI).

In this work, we have demonstrated parallel generation of entangled Bell states on an optical clock transition using a novel spin-echoed adiabatically resonant Rydberg pulse sequence. We characterize the Bell state fidelity to be $\mathcal{F} = 92.8(2.0)$\% and measure a 4.2(6) second array-averaged coherence lifetime. These results establish a firm foundation for future explorations of quantum-enhanced optical frequency metrology at the stability frontier. A crucial near-term goal in this direction is to extend these results from pairs to clusters of several-to-tens of atoms, and to identify and generate the particular optimal quantum states of these clusters for a clock stability measurement~\cite{khazali2016large, toth2014quantum}. Entangling operations both before and after the interrogation time may further enhance clock stability~\cite{davis2016approaching, hosten2016quantum, kaubruegger_variational_2019, kaubruegger2021quantum, schulte2020ramsey, colombo2021time}.

More broadly, we have demonstrated capabilities useful throughout quantum science. The combination of control and detection of highly coherent qubits, scalability to large system sizes, and entanglement generation are common requirements for a host of quantum applications. Particularly exciting goals include the generation of cluster states and, with the addition of local measurements, subsequent demonstration of one-way quantum computing~\cite{briegel2009measurement}. Explorations of the transverse field Ising model with finite range interactions are directly within reach. The combination of tweezer-based atom rearrangement and a 3d optical lattice further allows access to 2d Hubbard physics with entangled or even finite-range-interacting tunnelers. 

\section{Acknowledgements}
We acknowledge Jun Ye and his lab, in particular Dhruv Kedar, for the operation and provision of the silicon-crystalline-cavity-stabilized clock laser and helpful conversations. We thank Ivan Deutsch, Ana Maria Rey, Christian Sanner, and Alec Cao for fruitful discussions and feedback on the manuscript. This work was supported by ARO, AFOSR, DOE QSA, NSF QSEnSE, NSF Physics Frontier Center at JILA (1734006), and NIST. N.S. acknowledges support from the NRC research associateship program. W.J.E. acknowledges support from the NDSEG. M.J.M. was supported by the Laboratory Directed Research and Development program of Los Alamos National Laboratory under Project numbers 20190494ER and 20200015ER.
	
\section{Author Contributions}
The experiment was designed and built by N.S., A.W.Y., W.J.E., and A.M.K. N.S., A.W.Y., and W.J.E. operated the experiment and analyzed data. All authors contributed to the manuscript. 
	
\section{Author Information}
The authors declare no competing financial interests. Correspondence and requests for materials should be addressed to A.M.K. (adam.kaufman@colorado.edu). 
	
\section{Data Availability}
The experimental data presented in this manuscript are available from the corresponding authors upon request.

\bibliographystyle{naturemag}
\bibliography{theBib}

\renewcommand{\appendixname}{Methods}
	
\setcounter{equation}{0}
\setcounter{figure}{0}
\renewcommand{\theequation}{M\arabic{equation}}
\renewcommand{\thefigure}{M\arabic{figure}}

\clearpage

\setcounter{secnumdepth}{2}

\section*{Methods}

\subsection{Apparatus}
\label{methods:Apparatus}

\begin{figure}[b]
	\centering
	\includegraphics[width=0.45\textwidth]{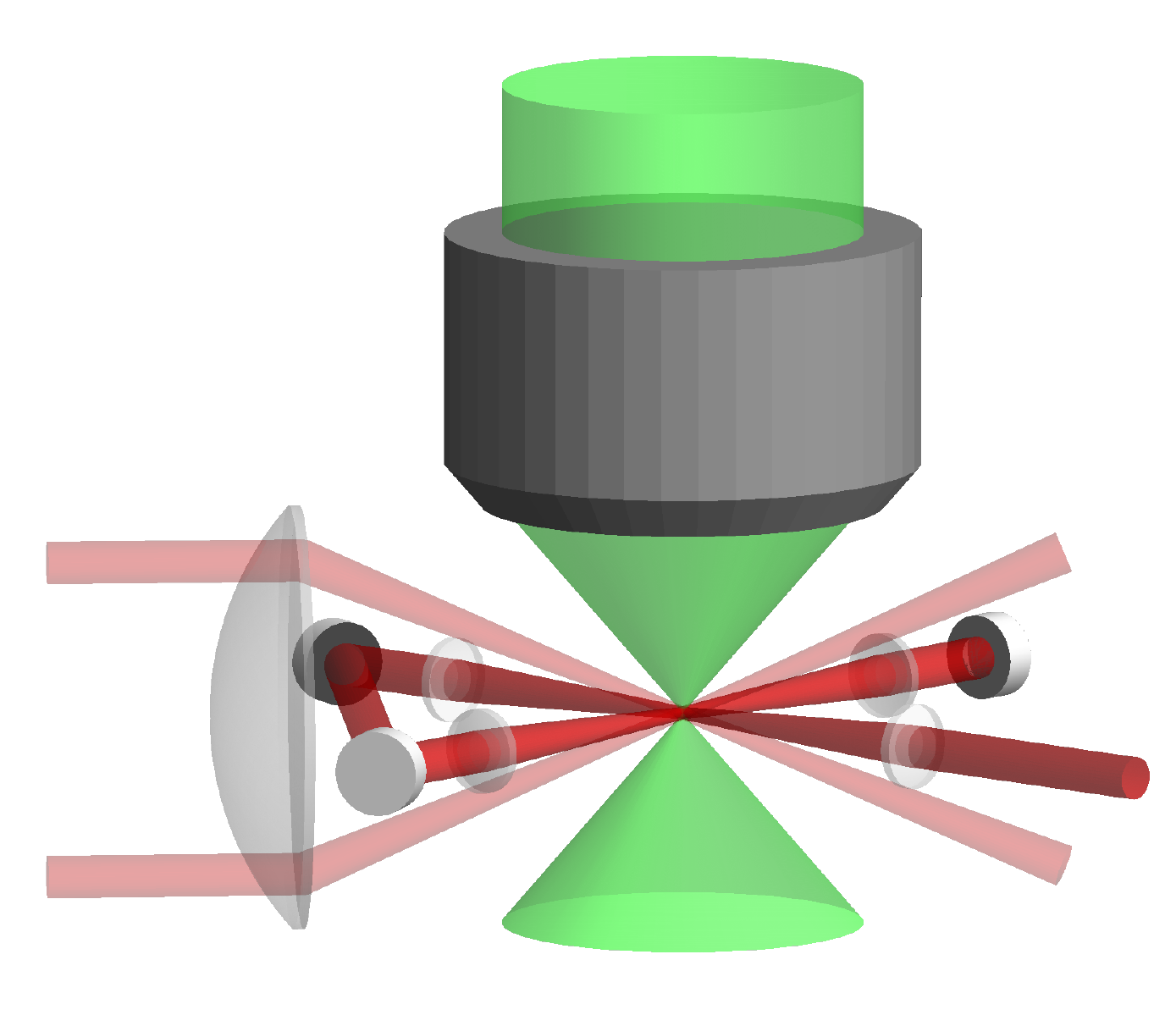} 
	\caption{\textbf{Trapping potentials of our apparatus.} Our experiments make use of three separate optical trapping potentials. After producing a cold atomic sample via standard MOT techniques, we load atoms into an array of 515 nm optical tweezers projected through a high-numerical-aperture objective. This forms a 2d array of atoms which are then transfered into a single plane of a 3d optical lattice which is formed by two separate optical systems. First, a 2d bowtie lattice is formed using a single beam redirected by a series of mirrors and lenses. A second, axial lattice is projected from the side by directing two path-length-matched beams into a single aspheric lens.}
	\label{fig:TrappingPotentials}
\end{figure}

Loading, cooling, and imaging of $^{88}$Sr atoms is described in \cite{norcia_microscopic_2018}. The rf source for generation of the 515 nm tweezer array is described in \cite{young_half_2020}. We use a 3d clock-magic optical lattice at 813 nm as the primary science potential for imaging, clock rotations, and Rydberg spectroscopy (see Fig. \ref{fig:TrappingPotentials} and SI for details). This potential enables efficient cooling and high-contrast clock rotations, as discussed below. 

\begin{figure}[h!]
	\centering
	\includegraphics{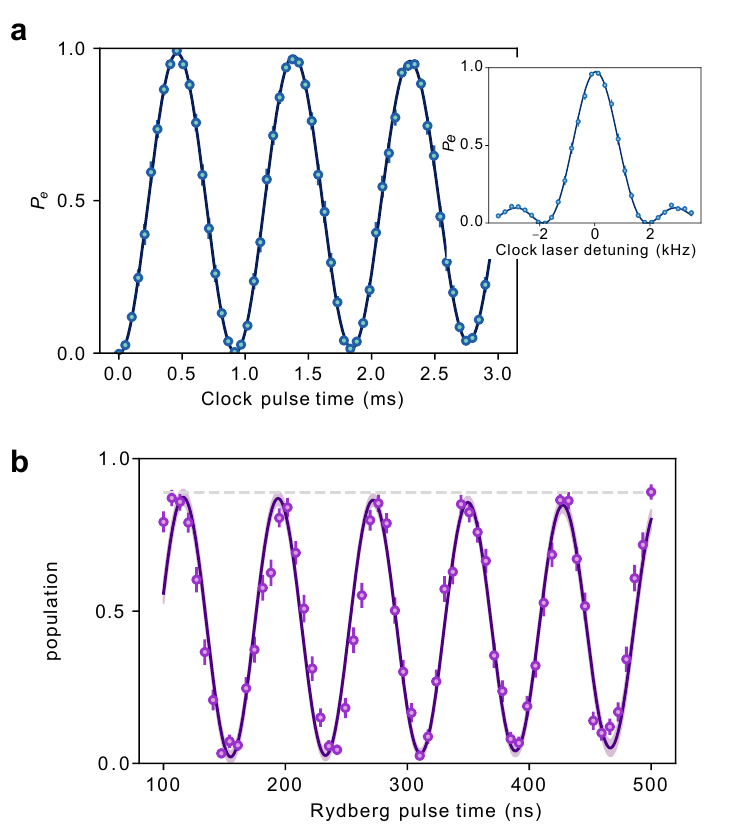}
	\caption{\textbf{Single atom control. a)} Tight atomic confinement and a 550 G magnetic field enable a clock Rabi frequency above 1 kHz with $\pi$-pulse fidelity well above 99\%; however, fidelity is limited to 99.41(57)\% in part by atomic heating (which also causes dephasing at later times, see SI). \textbf{Inset} shows the corresponding clock transition $\pi$-pulse spectrum. \textbf{b)} A UV laser near 317 nm drives a clock-Rydberg transition with Rabi with frequency $\Omega_r=2\pi\times$ 13 MHz. We detect the Rydberg state via loss of the clock state atoms. The horizontal dashed line represents an 89\% Rydberg detection fidelity estimated from branching ratios of intermediate triplet $S$ states (see text).} 
	\label{fig:clockandrydrabi}
\end{figure}

We open the $^{88}$Sr $^1$S$_0 \leftrightarrow$ $^3$P$_0$ optical clock transition by applying a vertically oriented magnetic field with magnitude up to 550 G. This is accomplished using our main (MOT) coils driven by a remotely-controlled Delta Electronika (SM30-200) supply. We drive the clock transition using approximately 13 mW of 698 nm laser light focused by a 150 mm cylindrical lens. The clock laser source is a series of three diodes injection-locked to light stabilized to a cryogenic silicon cavity~\cite{oelker2019demonstration}. The path from the reference laser to our experiment includes 60 m of fiber and approximately 2 m of free space which are noise-canceled, with the reference mirror attached to the main objective mount. In addition, there is approximately 1 m of propagation that includes the second injection locked diode which is not noise-canceled. Given the stability shown in our previous work with similar lengths of uncanceled free-space propagation~\cite{young_half_2020}, we do not expect the added phase noise from the uncanceled path length to limit clock pulse fidelity. As shown in figure~\ref{fig:clockandrydrabi}(a), we can drive the clock transition with a Rabi frequency in excess of 1 kHz and with a $\pi$-pulse fidelity of $99.41(57)\%$. This represents a significant increase over approximately $80\%$ fidelity shown in our previous work~\cite{young_half_2020}.

\begin{figure*}
	\centering
	\includegraphics{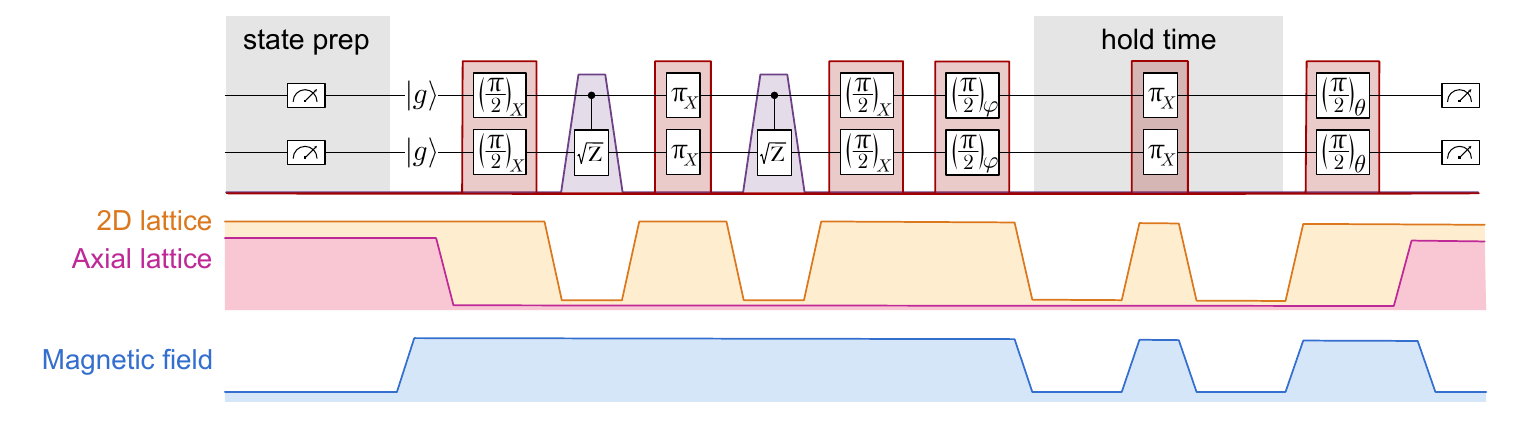}
	\caption{\textbf{Experimental Sequence.} We present a more detailed view of the experimental sequence that includes the clock laser pulses (red), the Rydberg laser pulses (purple), the 2d lattice (orange), the axial lattice (pink), and the z-component of the magnetic field (blue). Typical lattice ramping times are 3-10 ms, and we allow the magnetic field to ramp and settle for 100 ms. Although we can achieve clock Rabi frequencies as high as 1 kHz, we operate a factor of 10 slower to reduce the effect of magnetic field noise. As such, the total gate time from the first $\pi/2$-pulse through the second $\pi/2$-pulse is 22.2 ms, with 12 ms arising from the four 2d lattice ramps within the gate sequence.}
	\label{fig:exptsequence}
\end{figure*}

\begin{table*}
	\centering
	\begin{tabular}{|l|c|c|c|c|c|}
		\hline
		Parameter & Figure~\ref{fig:gate}d & Figure~\ref{fig:parity}a,b & Figure~\ref{fig:lifetime}a-e & Figure~\ref{fig:lifetime}a & Figure~\ref{fig:lifetime}f,g \\
		& & & blue, gray & purple & \\
		\hline 
		Rydberg detuning (initial, final) ($2\pi\times$ MHz) & (40,2)  & (22,2) & (22,5.8) & (40,4) & (40,8)  \\ 
		\hline 
		UV ramp time (ns) & 350 & 250 & 250 & 350 & 350  \\ 
		\hline 
		Rydberg Rabi frequency ($2\pi\times$ MHz) & 14 & 16 & 16 & 10 & 16  \\ 
		\hline 
		Resonant gate time (ns) & varied & 21 & 20 & 250 & 115  \\ 
		\hline
		Lattice states during UV pulses (2d, axial) ($E_r$) & (0,0) & (12,8) & (12,8) & (0,0) & (0,0)  \\ 
		\hline
		Magnetic field (during hold time if applicable) (G) & 550 & 55 & 55 (2.8) & 55 (55) & 55  \\ 
		\hline
	\end{tabular} 
	\caption{\textbf{Experimental parameters} Select parameters are tabulated for each dataset presented in the main text.}
\end{table*}

Our high-fidelity clock rotations are enabled by the improved spatial confinement afforded by a deep, power-efficient 3d lattice as opposed to shallow tweezers. Tight confinement enables resolved sideband cooling after transfer into the lattice, yielding a reduced average phonon number $\bar{n} = 0.04_{-0.04}^{+0.11}$ in the direction of the clock laser propagation. Tight confinement also manifests as a high 99 kHz trap frequency and correspondingly small Lamb-Dicke parameter $\eta = 0.22$. Both lower temperature and smaller Lamb-Dicke parameter result in reduced motional dephasing of Rabi oscillations, yielding an expected (ideal) $\pi$-pulse fidelity of 99.98$_{-.09}^{+.02}$\%. In typical experiments, however, we observe some heating which limits the clock $\pi$-pulse fidelity to a maximum of 99.79\% (see SI), which may be compared to our observed clock $\pi$-pulse fidelity of 99.41(57)\%.

We connect the clock state to a Rydberg state using a UV laser near 317 nm with approximately 1 W of power. The UV laser system first produces 634 nm light via single-pass sum frequency generation of 1066 nm and 1569 nm. A final stage of second harmonic generation inside a resonant cavity converts 634 nm light into 317 nm light (see SI for more details). We drive the transition from the clock state to the 5s40d $^3$D$_1$ $m_j$=0 Rydberg state with Rabi frequencies up to $\Omega_r = 2\pi\times$ 18 MHz. We choose a $D$ state rather than an $S$ state due to its greater dipole matrix element, which directly increases the entangling energy in our adiabatically resonant pulse scheme, while the reduced van-der-Waals $C_6$ interaction coefficient is still sufficient to ensure blockade at the 1.1 $\mu$m intradoublet separation while enabling denser doublet spacings. We observe Rydberg spectra and Rabi oscillations through the loss of clock state atoms (see Fig.~\ref{fig:clockandrydrabi}b). Perhaps surprisingly, the observed contrast is consistent with the 1/9 branching ratio of intermediate triplet S states back into the clock state versus to all 5s5p $^3$P$_J$.

\subsection{Experimental Sequence}

\begin{figure*}
	\includegraphics[width=0.75\textwidth]{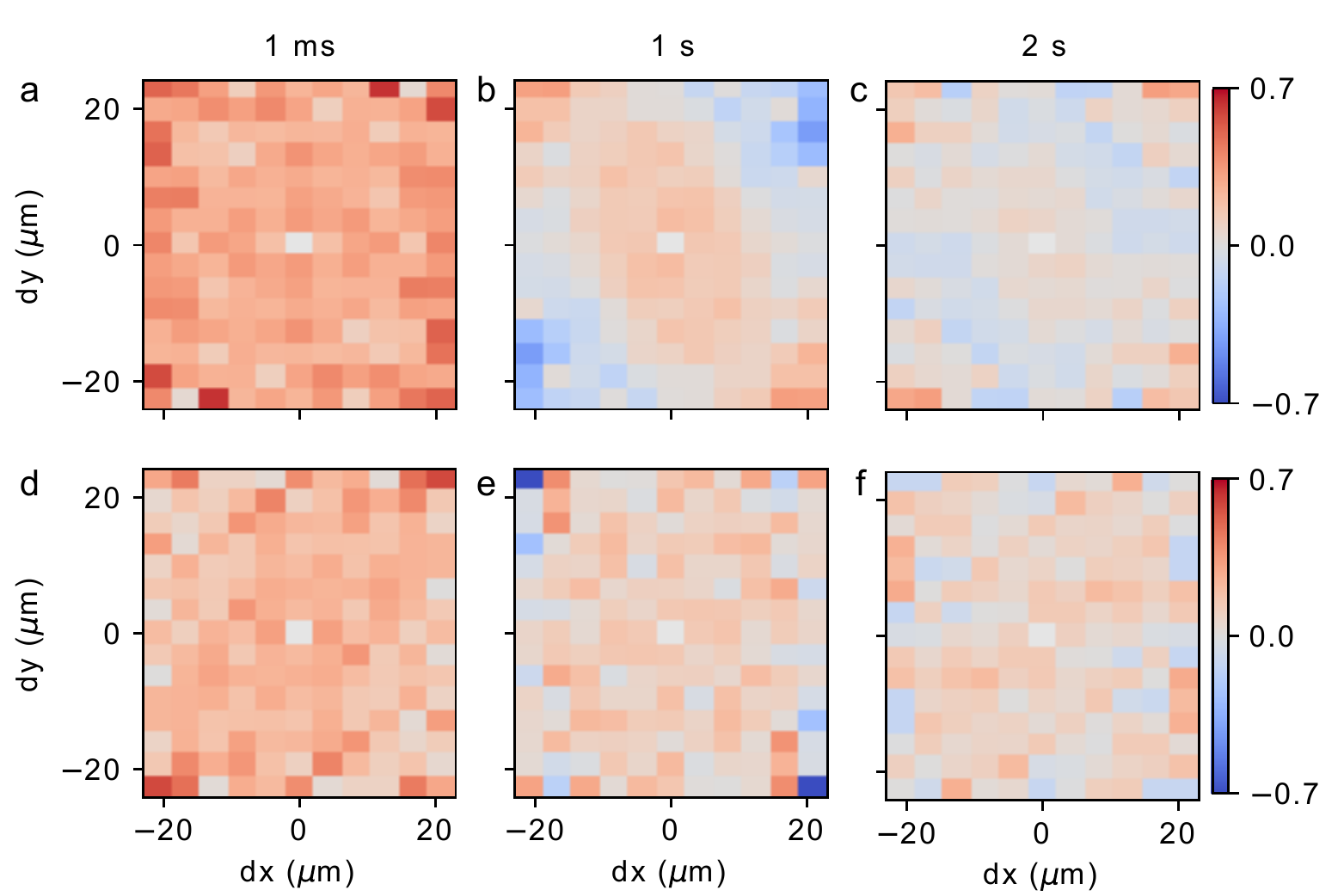}
	\caption{\textbf{Removing inhomogeneous lightshifts across the atom array.} By increasing lattice laser detuning from the magic condition, we more clearly reveal the effects of a non-magic-lattice inhomogeneity in spatially resolved parity-parity correlation measurements at short times (\textbf{a},\textbf{d}), after 1 second (\textbf{b},\textbf{e}), and after 2 seconds (\textbf{c},\textbf{f}). The location of each pixel corresponds to a displacement vector between fully filled doublets, while the color corresponds to the product of the parities of each doublet with that displacement vector. \textbf{a-c)} Even in a shallow lattice condition, significant residual inhomogeneous lightshifts result in a variation of the parity-parity correlation across the array, reducing the array-averaged parity-parity correlator. \textbf{d-f)} Introducing a $\pi$-pulse half way through the hold time spin-echoes away the inhomogeneity, increasing the array-averaged parity-parity correlator.}
	\label{fig:spatiallyresolvedcorrlator}
\end{figure*}

Our experiments begin by preparing a MOT, loading and cooling single atoms into a 515 nm wavelength tweezer array, transferring those atoms into a single plane of a 3d optical lattice at 813 nm, imaging the location of those atoms, and cooling the atoms, resulting in $>$90\% of atoms in their 3d motional ground state (labeled state prep in Fig.~\ref{fig:exptsequence}). This process has been largely described in our previous work~\cite{norcia_microscopic_2018,young_half_2020}, except for the transfer into and cooling in the 3d optical lattice. The loss due to transfer from tweezers into well-aligned lattice sites is negligible; however, the global alignment of the tweezer array to the underlying lattice is not perfectly commensurate, resulting in `slips' in the spacing of atoms in the lattice, which causes variable imaging performance. Furthermore, we find that the relative spatial phase between the tweezer array and lattice drifts by a lattice-site over the course of several hours. Improving the alignment and stability of this system will be the subject of future work. 

Ground state cooling proceeds similarly to cooling in 813 nm tweezers described in previous work~\cite{norcia_microscopic_2018,young_half_2020}. Achieving a magic angle between the applied magnetic field and optical polarization enables efficient cooling to the 3d motional ground state. This does require, however, that the axial lattice be vertically polarized, which makes it less power efficient and leads to the formation of an out-of-phase horizontally polarized lattice as well. Nevertheless, we typically achieve 3d ground state fractions above 90\% although the exact performance can vary day-to-day.

Having prepared an ensemble of ground state atoms in known positions in the lattice, we are ready to perform our experiments. In order to open the clock transition, we apply a large magnetic field. While we can apply a field of up to 550 G to yield a clock Rabi frequency above 1 kHz, we typically apply a 55 G field, resulting in a 100 Hz clock Rabi frequency. The use of lower fields slows the gate and does not substantially affect gate fidelity, but it does reduce magnetic field noise which otherwise limits parity oscillation decay times to several tens of milliseconds due to the fluctuating quadratic Zeeman shift of the clock state (see SI). 

Throughout the experimental sequence, we hold the axial lattice low, at approximately 8 E$_r$, while we ramp the 2d lattice high, to 420 E$_r$, for all clock pulses, and low, to 12 E$_r$, for all Rydberg pulses, and 37 E$_r$ for all hold times. Note that E$_r$ refers to the recoil energy of a single 813 nm photon. Ramping the 2d lattice high provides strong confinement for a low Lamb-Dicke parameter $\eta = $0.22, which enables high-fidelity clock pulses, while ramping the 2d lattice low reduces the total lattice-induced lightshift on the Rydberg line to $\sim$200 kHz. This is sufficiently low that lattice inhomogeneity does not affect gate fidelity, while leaving the lattice on provides atomic confinement. Experiments with the lattice switched off during the UV pulses can result in atomic heating, increased loss, and lower gate fidelity.

Finally, in the relevant experiments with a variable hold time, we perform a spin-echo clock $\pi$-pulse to remove dephasing due to spatial inhomogeneity of the clock frequency. This inhomogeneity arises from the lattices that confine the atoms --- since their optical frequencies are separated by 160 MHz, both lattices cannot simultaneously be at the clock-magic wavelength. Combined with the $\sim 50$\% peak-to-peak inhomogeneity of the combined lattice depth across the atom array, this yields a lightshift inhomogeneity of $\sim0.1$ Hz. This causes a significant reduction in the various correlation decay times measured in figure~\ref{fig:lifetime}. By introducing a $\pi$-pulse in the middle of the hold time, we greatly reduce the effects of inhomogeneity across the array, as can be seen in the spatially resolved parity-parity correlation measurements of figure~\ref{fig:spatiallyresolvedcorrlator}. 

\renewcommand{\tocname}{Supplementary Information}
\renewcommand{\appendixname}{Supplement}
	
\setcounter{equation}{0}
\setcounter{figure}{0}

\renewcommand{\thefigure}{S\arabic{figure}}

\incltocpage
\clearpage

\tableofcontents
\appendix
\setcounter{secnumdepth}{2}
\numberwithin{equation}{section}

\section{Experiment}

\subsection{Apparatus}
\label{SI:apparatus}

\subsubsection{Optical lattice}

Our 3d optical lattice at 813 nm is composed of two parts: a 2d optical lattice and an axial lattice. The 2d optical lattice is a vertically (along tweezer propagation direction) polarized bowtie lattice, defined by two folding mirrors, a retroreflecting mirror, and four lenses all mounted on the same macor plate that holds the main objective for imaging and tweezer array projection. This helps to ensure that the relative position of the 2d lattice and tweezer array remains stable. The lenses are in two 4f telescope pairs so that the beam is collimated at all mirrors and is focused to a 55 $\mu$m waist where it overlaps with the atom array. We send approximately 600 mW of 813 nm light to achieve 420 $E_r$ trap depth with radial trapping frequencies of $99$ kHz. To provide strong confinement in all three dimensions, we further incorporate an axial lattice at 813 nm, detuned by 160 MHz from the bowtie lattice, utilizing a similar design to our previous 515 nm axial lattice~\cite{young_half_2020}. We send approximately 300 mW of light down each of two parallel, vertically displaced beams incident on a 30 mm aspheric lens which then intersect at a 36 degree angle. Both lattice beams are vertically polarized before the final lens. We find a $42$ kHz axial trap frequency in a 400 $E_r$ deep potential. To reduce slow drift of the axial lattice phase with respect to the plane of the atom array, the path lengths of the two beams are matched to better than 1 mm. Nevertheless, we must match the axial lattice phase to the plane of the atoms every few hours, which we accomplish by minimizing atom loss from parametric modulation of the trap depth at the trap frequency. This procedure is outlined in \cite{young_half_2020}, SI. A2. 

\subsubsection{UV laser system}

\begin{figure*}
	\centering
	\includegraphics[width=\textwidth]{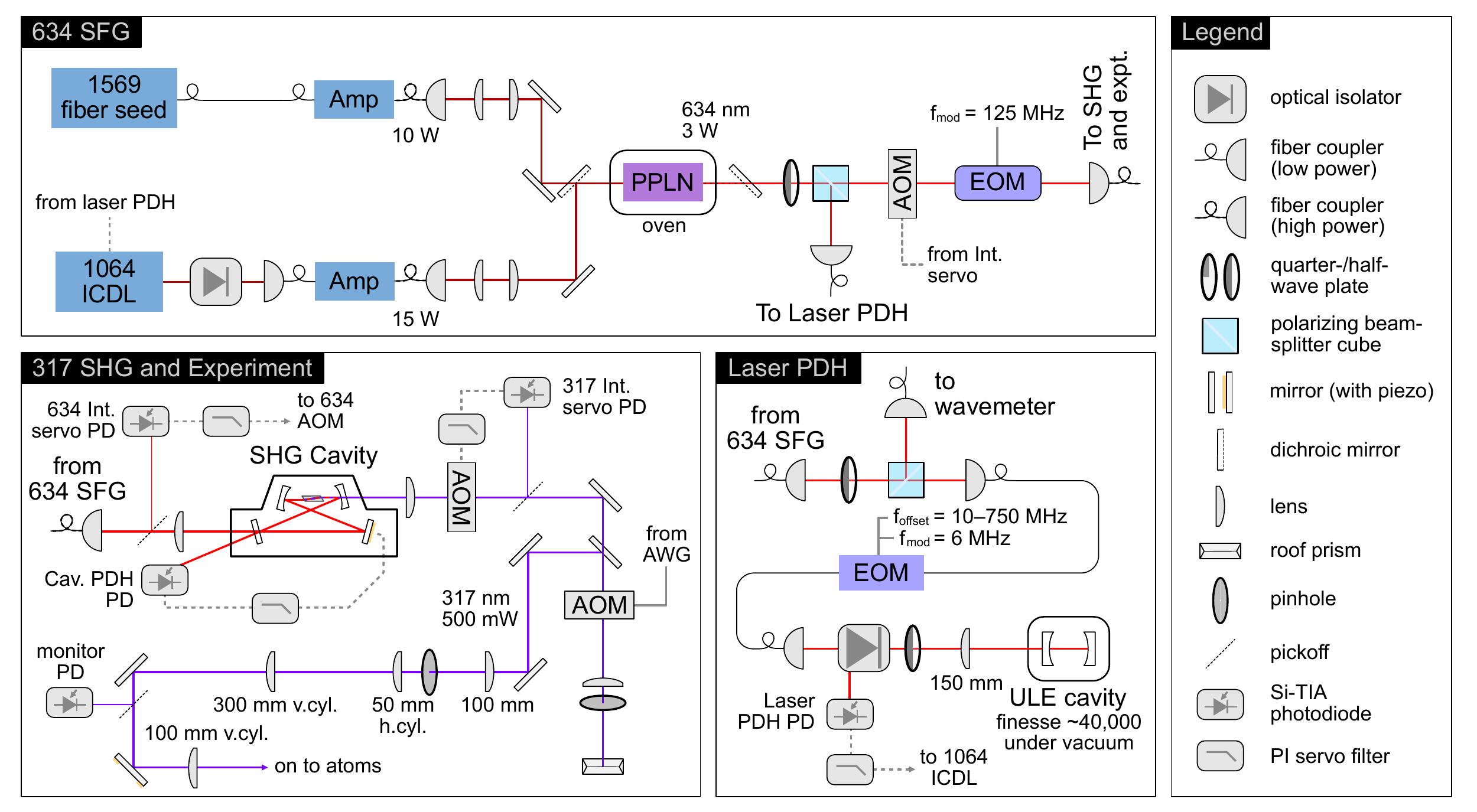}
	\caption{\textbf{UV laser system.} A schematic of our system for generating over 1 Watt of laser power at 317 nm is shown. The system is split into three modules. The first uses two IR seed lasers and fiber amplifiers to perform free space sum frequency generation (SFG) of 634 nm light. Some of this light is sent to a stabilization system which feeds back on the 1064 nm seed laser to stabilize the 634 nm laser frequency. A high power fiber takes the primary output of the 634 nm laser and directs it into a home built resonant second harmonic generation (SHG) cavity to produce 317 nm light with 42\% efficiency. The 317 nm output of the SHG cavity is intensity stabilized, pulse shaped, and mode cleaned before being directed into the experiment.}
	\label{fig:UVsystem}
\end{figure*}

Here we provide details of our UV laser system, which relies on one stage of frequency summing followed by a stage of frequency doubling. These nonlinear optical techniques require high optical power for efficient conversion. The combination of these specifications is well matched to the capabilities of fiber lasers and fiber amplifiers to produce $\sim 10$ W of narrow linewidth, stable power in the near-infrared. Conveniently, the half-frequency of the required 317 nm light is at 634 nm which can be produced from the sum of 1066 nm and 1565 nm photons, both of which are in frequency bands of commercially available fiber amplifiers. Following the design of similar UV lasers in the NIST ion storage group,~\cite{wilson2011750} we seed a 1569 nm amplifier (NKT K532-015-120) with 20 mW from a passively stable fiber laser (NKT K822-175-112). For tunability of the UV wavelength, we seed the 1064 amplifier (NKT K532-005-130) with 12 mW from a home-built interference cavity diode laser (Sacher Lasertechnik AR coated diode 210210600601, Semrock MaxLine filter LL01-1075-12.5x12.5). 

The outputs of the two fiber amplifiers are passed through beam shaping optics and sent into an MgO:PPLN waveguide (Covesion MSFG637-0.5-40, 0.5(T)x10(W)x40(L) mm) which is temperature stabilized at 130$^\circ$C in an oven (Covesion PV40). This converts 10 W of 1569 nm light and 15 W of 1064 nm light into as much as 9.5 W of 634 nm light. A few mW are picked off and sent to a wavemeter (Toptica Photonics HF-ANGSTROM WS/6-US) and a stable ULE cavity (Stable Laser Systems SLS-6010-1-4bore) for slow frequency tunability and frequency stabilization via a PDH lock on an offset sideband. The remaining light is passed through an AOM (IntraAction AOM-802AF1) in a noise eater configuration for intensity stabilization followed by an EOM (Qubig PM8-VIS\_125+W+X5P3+TC) which introduces 125 MHz sidebands for PDH locking of the SHG cavity. This cavity is home-built based on designs from the ion storage group at NIST, Boulder~\cite{wilson2011750}. The cavity mirrors are sourced from Laseroptik GmbH (three mirrors with $R>99.9$\% and one incoupler with $R=98.4(2)$\% at 633 nm and all mirrors $R<5$\% at 317 nm). For second harmonic generation, the cavity includes a BBO crystal (Newlight Photonics, 4x4x10 mm, $\theta=37.9^\circ$, $\phi=0^\circ$, Brewster cut ends). To extend the crystal lifetime, we feed low flow-rate oxygen into the crystal housing, creating an oxygen rich atmosphere. The SHG cavity converts 634 nm light into 317 nm light with 42\% efficiency, in line with expectations given the nonlinear coefficient of BBO and achievable circulating power. UV damage of the crystal has only required translation and realignment of the cavity once over the course of a year of operation. 

The UV output of the SHG cavity passes through a fused silica AOM (IntraAction ASM-802B8) in a noise eater configuration to stabilize the intensity of the UV before passing through another fused silica AOM (IntraAction ASD-1802LA34.317) in double pass for fast frequency tunability and pulse shaping. Importantly, because this AOM only has high diffraction efficiency for S polarization, we retro-reflect with a roof-prism (Thorlabs PS610) with a small upwards displacement so that the double-pass beam may be isolated and directed to the experiment; this results in a double-pass efficiency of 75\%. The beam is focused through a pinhole (50 $\mu$m diameter, 55\% efficiency) to remove transverse structure in the laser mode typical from walkoff in the SHG crystal. The expanding beam is then collimated into an elliptical beam (240 $\mu$m x 1,870 $\mu$m $1/e^2$ intensity radii) and passed through a variable iris. Finally, the large vertical axis is focused onto the atom array via a 100 mm cylindrical lens, providing a vertical waist as small as 5 $\mu$m.

\subsubsection{Imaging}

Our experiments involve two images. The first, during state preparation, images ground state atoms to identify empty, half-filled, and fully-filled doublets. A second image concludes each experiment and identifies atoms in a particular state relevant to the experiment. Imaging involves collection and measurement of $\sim 15-25$ photons per atom in 200 ms. Photons are scattered from intensity-balanced, counter-propagating 461 nm laser beams, detuned 689 MHz from the $^1$S$_0$ $\leftrightarrow$ $^1$P$_1$ resonance and 20 MHz from each other. Simultaneous resolved sideband cooling in both radial and axial directions counterbalances recoil heating. 

Two back-to-back images can achieve combined loss and infidelity below 1\%. However, typical imaging calibration conditions --- which involve lattice loading slips, lattice depth ramps, and increased hold time --- yield combined loss and infidelity between 2\% and 4\%. Over the course of several hours, the imaging loss has been observed to fluctuate by $\sim$1\%.

Only ground state atoms are directly observed, but the application of a ground state blowaway beam at 461 nm and repump beams at 679 nm and 707 nm (resonantly driving $^1$S$_0$ $\leftrightarrow$ $^1$P$_1$, $^3$P$_0$ $\leftrightarrow$ $^3$S$_1$, and $^3$P$_2$ $\leftrightarrow$ $^3$S$_1$ respectively) further enables imaging of $^3$P$_0$ and $^3$P$_2$ state atoms. A blowaway pulse immediately following the conclusion of the experiment sequence removes ground state atoms. During a delay when the magnetic fields settle over $\sim100$ ms, clock state atoms may Raman scatter into $^3$P$_1$ (and from there decay to $^1$S$_0$) or $^3$P$_2$. Nevertheless, we recover the initial population in the clock state (assuming no initial population in $^3$P$_2$) by application of both repump lasers and subsequent imaging. We note that some experiments, e.g. Rydberg loss measurements, result in significant population in $^3$P$_2$, so for these data the 707 nm repump removes $^3$P$_2$ population before the ground state blowaway. 

\subsection{Atomic lifetimes and resulting Bell state coherence time expectations}
\begin{figure}[h!]
	\centering
	\includegraphics[width=0.85\columnwidth]{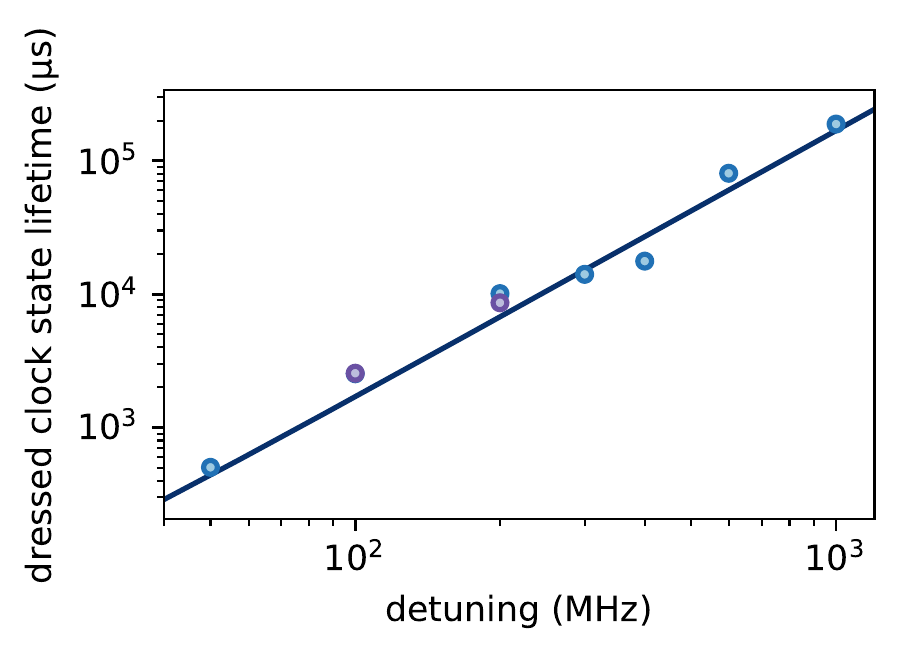}
	\caption{\textbf{Rydberg lifetime.} We measure the lifetime of clock state atoms trapped in a full depth lattice and dressed by the Rydberg coupling with Rabi frequency $\Omega_r = 2\pi\times$ 15 MHz and detuning $\Delta$ (negative detuning in blue, positive detuning in purple). The dependence of these lifetimes on detuning then provides the lifetime of the 5s40d $^3$D$_1$ Rydberg state, $\tau_r = 10.9(4)$ $\mu$s.}
	\label{fig:rydberglifetime}
\end{figure}
\begin{figure}[h!]
	\centering
	\includegraphics[width=0.75\columnwidth]{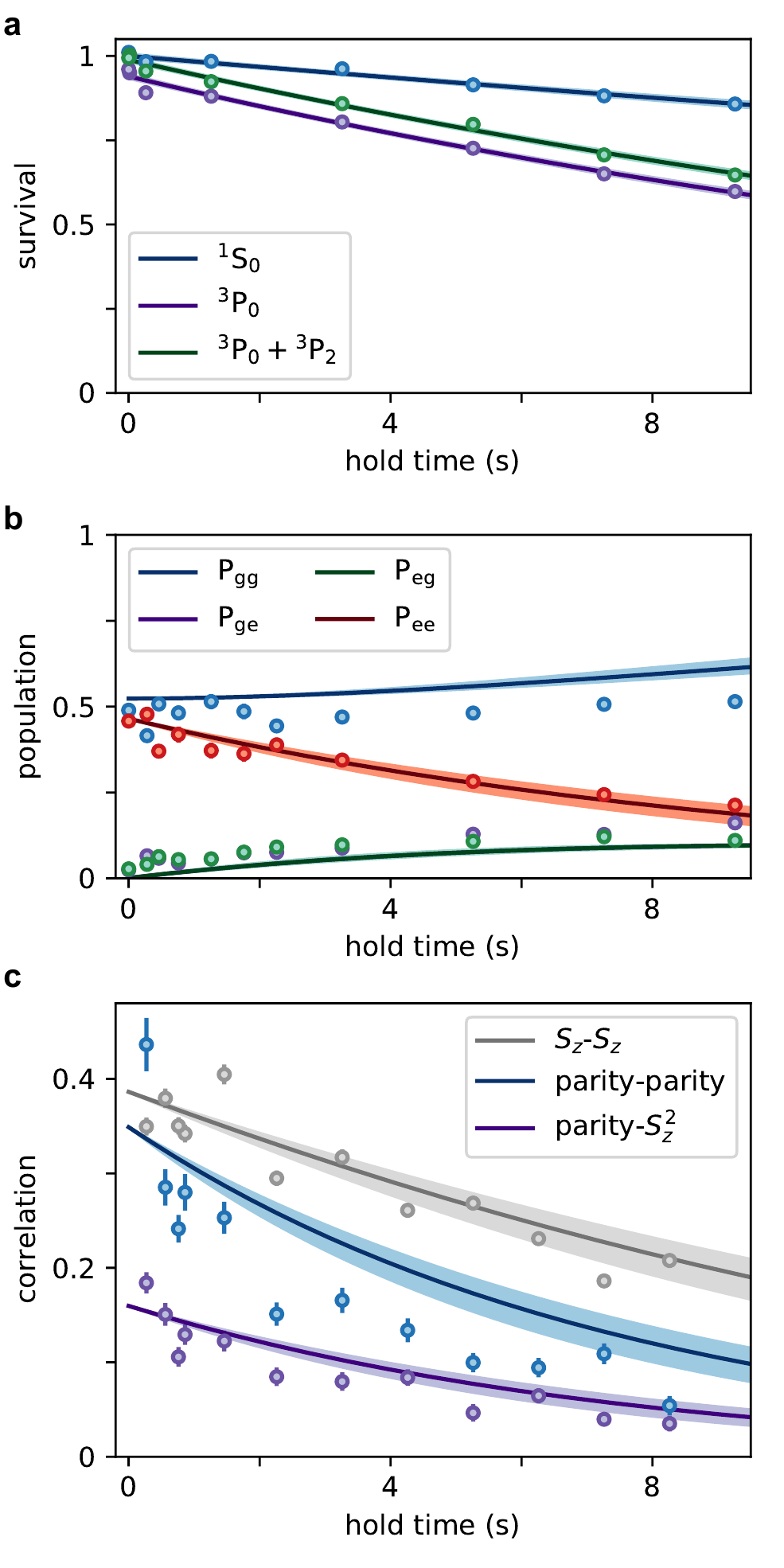}
	\caption{\textbf{Population lifetimes and inferred dynamics. a)} We measure the lifetime of atoms in either the ground state or clock state. For clock state atoms, we measure the population that remains either only in the clock state, or in both the clock and $^3$P$_2$ states. Exponential fits (curves) provide single atom decay rates. \textbf{b)} We measure two-qubit population dynamics of Bell states, primarily observing the decay of $P_{ee}$ into either $P_{ge}$ or $P_{eg}$ that further decay into $P_{gg}$ resulting in an increasing population in that state. A model using an estimation of the experimentally realized two-atom density matrix and single atom decay and loss processes as measured in panel \textbf{a} predicts the time dynamics of Bell state populations with no adjustable parameters. \textbf{c)} The same model offers a prediction for correlation decay from only single-atom decay and loss processes. It is presented on top of the same data from the main text figure~\ref{fig:lifetime}b,c.}
	\label{fig:atomiclifetimes}
\end{figure}

The loss or decay of atoms out of particular states impacts gate fidelity and the lifetime of resulting Bell states. First, we measure the radiative lifetime of the 5s40d $^3$D$_1$ Rydberg state, via the detuning dependence of dressed clock state lifetimes. We detune the Rydberg coupling of Rabi frequency $\Omega_r = 2\pi\times$ 15 MHz into the weakly-dressed limit, so that the clock-like eigenstate contains a small Rydberg state component, $P_r \approx \Omega_r^2/4\Delta^2$. This component yields loss $\tilde{\Gamma} = P_r\Gamma_r$ where $\Gamma_r$ is the Rydberg state lifetime, which includes radiative loss to low lying states and scattering by black-body radiation into nearby Rydberg states. We note that anti-trapping of the Rydberg state does not directly contribute to loss of the dressed clock state. However, the strong anti-trapping at full lattice depth of intermediate Rydberg states after a decay or scattering event efficiently prevents subsequent decay of the atom into the clock state, which would otherwise cause a systematic error. In figure~\ref{fig:rydberglifetime}, we observe the dressed clock state lifetime over a range of detunings. Given the calibrated Rydberg Rabi frequency, the dependence of lifetime on detuning provides the Rydberg state radiative lifetime, $\tau_r = 10.9(4)$ $\mu$s. We compare this data with separate measurements of the Rydberg lifetime that are directly sensitive to lattice anti-trapping. Two separated Rydberg $\pi$-pulses are applied to an ensemble of clock state atoms, and the decay of the Rydberg state appears as a loss of clock state atoms. Below a total lattice depth of $\sim50$ $E_r$, the observed Rydberg lifetime is consistent with 11 $\mu$s. Given the consistency of these observations, we use the dressed clock state lifetime detuning dependence to estimate the Rydberg state lifetime. This is incorporated into master equation simulations of the gate dynamics and Bell state fidelity (see section~\ref{SI:mastereqn}).

Next, we measure single-atom qubit state loss/decay rates in the absence of coupling to the Rydberg state, as these rates directly impact single atom coherence and thus the Bell state coherence time. In figure~\ref{fig:atomiclifetimes}a, we prepare the atomic ensemble in either the ground state ($^1$S$_0$) or the excited state ($^3$P$_0$), and measure the population remaining in that state after a variable hold time. The ground state is lost with an exponential timescale of 60.5(5.2) s, and the excited state either decays or is lost with an exponential timescale of 20.2(8) s. The average of the single atom $\ket{g}$ and $\ket{e}$ population loss/decay rates provides an upper bound of $\tau_{pop} = 30.3(1.1)$ s on the measured single atom coherence lifetime, $2\tau_{S_zS_z} = 24.4(9)$ s. The difference between these observations is likely caused by a lossless decoherence process such as the Rayleigh scattering of blackbody radiation, which occurs at a rate $\Gamma_{BBR}^0 = 3.45(22) \times 10^{-3}$ s$^{-1}$~\cite{dorscher_lattice-induced_2018}. Accounting for blackbody radiation Rayleigh scattering leaves an unexplained lossless decoherence process with an exponential lifetime of 220(100) s. 

The Bell state coherence time is upper bounded by the average of the lifetimes of $\ket{gg}$ and $\ket{ee}$, which are simply obtained as half the lifetime of the single atom qubit states, assuming the absence of correlated loss or decay processes during the hold time. We have directly searched for such processes and not found them. To search for possible correlated loss processes in Bell states and to provide a timescale for population decay for comparison with the parity contrast decay of figure~\ref{fig:lifetime}a, we prepare an array of Bell states and measure the resulting two-qubit state populations after a variable hold time (see Fig.~\ref{fig:atomiclifetimes}b). We observe the decay of $P_{ee}$ into either $P_{ge}$ or $P_{eg}$ with an exponential timescale of 12.3(1.0) s, which is consistent with a doubling of the single atom $\ket{e}$ state decay rate and the absence of correlated loss or decay processes of clock state atoms.

We validate this consistency through a simple master equation model, where we construct the initial density matrix for the simulation from Bell state measurements just before the hold time in our lifetime measurements. For additional comparisons with the $S_z$-$S_z$ and parity-$S_z$ correlators, we also construct an initial density matrix for simulation from the measured total spin projection of the ensemble of unentangled single-atoms. The model Hamiltonian is simply the identity operator, and there are two jump operators associated with atom loss and clock state decay, with exponential time constants $\tau_{\text{trap}}$ and $\tau_{\text{decay}}$, respectively. Based on our measured ground and clock state lifetimes  $\tau_g$ and $\tau_e$, we set $\tau_{\text{trap}}$ = $\tau_g$ = 60.5 s and $1/\tau_{\text{decay}} = 1/\tau_e - 1/\tau_g = 1 / 30.4 \ \text{s}^{-1}$. The predicted two-qubit population dynamics, shown in figure~\ref{fig:atomiclifetimes}b, show reasonable agreement with the data with no adjustable parameters. 

The comparison of $\tau_{pop}$ and $2\tau_{S_zS_z}$ together with the absence of evidence for correlated loss or decay processes suggests that the $S_z$-$S_z$ correlation lifetime of 12.2(5) s is largely, though not completely, explained by measured single atom loss and decay processes. This timescale, which is half the single-atom atomic coherence time, is significant primarily as an expectation for the Bell state coherence lifetime, which at $\tau_{bc} = 4.2(6)$ s, is significantly shorter. We further explore the discrepancy between observed correlation decay times and the expectation from measured single atom loss and decay processes in figure~\ref{fig:atomiclifetimes}c. The same master equation model yields a prediction for these observables with no free parameters. Of particular note, the model curves all decay to zero correlation at long times in disagreement with resolved offsets in exponential fits to data. The data qualitatively appear to more closely match the late time model curve decay while decaying faster at early times, suggesting the presence of two decay timescales rather than a single faster decay timescale. While this could be the result of unexplained correlated dephasing processes, we hypothesize that it is due to technical noise in the experiment relating to switching between the lattice and magnetic field conditions used in the gate, and during the long hold times when measuring lifetimes. 

To provide a conservative estimate of the lifetime in the presence of this two-timescale decay, and other potential effects like correlated loss or decay with multiple timescales, we allow for the possibility of an offset in our fits to the data. This is done via a goodness-of-fit comparison that accounts for the change in the fit degrees of freedom. The $S_z$-$S_z$ correlation decay is well described by decay without an offset. The hypothesis that the parity-parity correlation data is better fit with without an offset is weakly disfavored with a p-value of 0.18. Although the parity-$S_z^2$ correlation decay has a corresponding p-value of 0.43, indicating no clear preference in the fit form, we understand this curve to be composed of the same information contained in the parity-parity and $S_z$-$S_z$ correlations, and since the parity-parity correlation fits with an offset, we include an offset in our fit of the parity-$S_z^2$ correlation. These fits appear in figure~\ref{fig:lifetime}b,c, and are the basis for the lifetimes quoted above and in the main text. However, it is worth pointing out that the late-time behavior of our data is in agreement with single-atom measurements. In fact, if we fit the data without an offset, as is physically motivated by our understanding of the correlation functions, we find that the parity-parity correlation decays with an exponential time constant of $\tilde{\tau}_{\Pi\Pi}  = $ 5.0(3) s and the parity-$S_z^2$ correlation decays with an exponential time constant of $\tilde{\tau}_{\Pi S_z^2} = $ 5.3(4) s. From these we can infer Bell state coherence times $\tilde{\tau}_{bc}^{\Pi\Pi} = $ 10.0(6) s and $\tilde{\tau}_{bc}^{\Pi S_z^2} = $ 9.5(1.4) s, for a combined inferred Bell state coherence time of $\tilde{\tau}_{bc} = $ 9.9(5) s, which largely removes the discrepancy between the observed Bell state coherence time and our expectations based on single atom measurements.

\section{SPAM correction}

An atom is detected through its fluorescence signal on a camera when the number of photons detected on a pre-defined region of the sensor exceeds a threshold. This process works so long as the photon number from an atom fluorescing is sufficiently well distinguished from the background photon number, and the process of scattering photons off an atom of interest does not heat the atom out of its trap. This statement points to the three main error syndromes of fluorescence detection, vacancy-atom infidelity, atom-vacancy infidelity, and loss:
\begin{enumerate}
    \item A vacancy (a location without an atom) is misidentified as a location with an atom, with probability $p_{iva}$, 
    \item an atom is misidentified as a vacancy, with probability $p_{iav}$, and
    \item an atom is lost during imaging, with probability $p_{loss}$
\end{enumerate}

All experiments we run involve initial state preparation, a first image, further state preparation, the experimental sequence, and a final image. All resulting data appear in the site-resolved loss of atoms between the first and second image. Functionally, these error processes result in either an atom detected when there would have been a vacancy or the reverse, so we define the error rates $E_{va} = p_{iva}$ and $E_{av} = p_{iav}+p_{loss}$. In order to correct for these errors, we use imaging calibration datasets to extract the probabilities $E_{av}$ and $E_{va}$. These datasets involve running the full experimental sequence except for clock, UV, and blowaway pulses. We denote the resulting calibration and ultimate correction of errors as SPAM rather than purely measurement error correction because various hold times and lattice ramps can also contribute to the calibrated and corrected loss. Given these two probabilities, we construct a measurement process matrix which represents the conversion of true experimental events into detected experimental observations. The inverse of the measurement process matrix is then applied to measured datasets to extract, with appropriate added uncertainty, our estimate of the true reality of our experiment's results. More explicitly, 
\begin{equation}
    \mathbf{m} = M \cdot \mathbf{N}
\end{equation}
where $\mathbf{N} = \{N_{aa}, N_{av}, N_{va}, N_{vv}\}$ is the column vector of the number of true events with an atom present at the time of both images, with an atom present only at the first image and a vacancy at the second image, with a vacancy at the first image and an atom at the second image, and with a vacancy at both images. Note $N_{va}=0$ unless atoms are not pinned between the two images. $\mathbf{m} = \{m_{aa}, m_{av}, m_{va}, m_{vv}\}$ is likewise the column vector of the number of \emph{measured} events of the four possible types. These two column vectors are connected by the measurement process matrix, given by

\begin{widetext}
\begin{equation}
    M = \left(
  \begin{tabular}{cccc}
$1-2E_{av}$ & $E_{va}$ & $E_{va}$ & 0 \\ 
$E_{av}$ & $1-E_{av}-E_{va}$ & 0 & $E_{va}$ \\ 
$E_{av}$ & 0 & $1-E_{av}-E_{va}$ & $E_{va}$ \\ 
0 & $E_{av}$ & $E_{av}$ & $1-2E_{va}$ \\ 
\end{tabular}
    \right)
\label{eqn:measurementmatrix}
\end{equation}
\end{widetext}
This form assumes that the error probabilities are small and so neglects second order processes, and that error probabilities are uniform across the array and between the two images. 

\begin{figure}[h]
    \centering
    \includegraphics[width=\columnwidth]{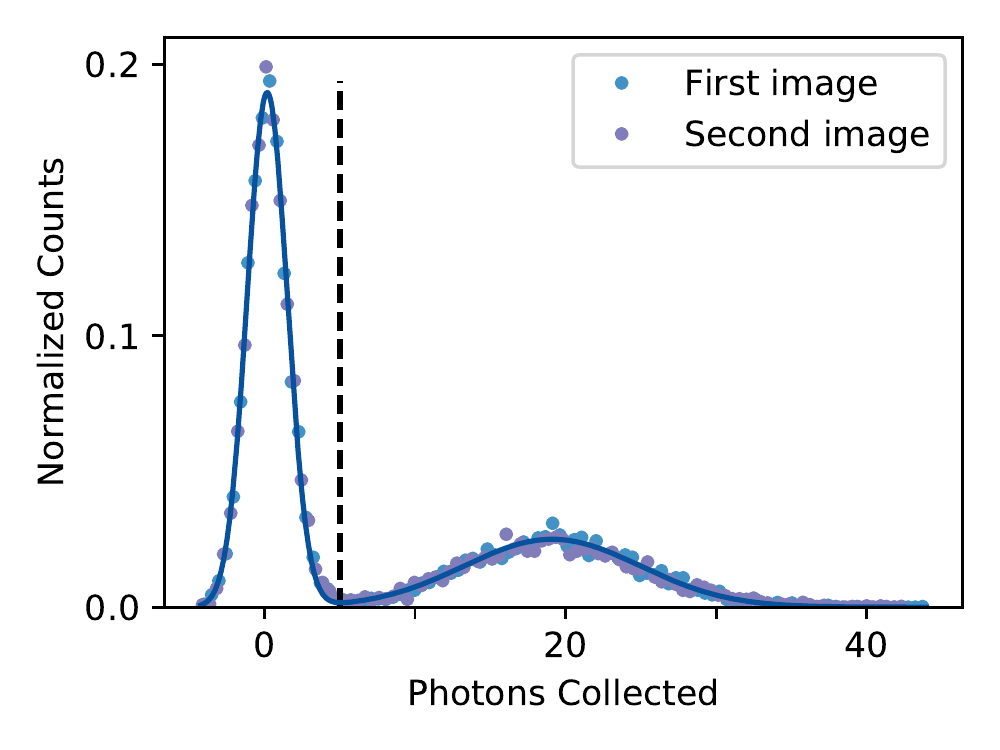}
    \caption{\textbf{Imaging Histogram.} We present imaging calibration histograms for the Bell state fidelity data presented in figure~\ref{fig:parity}. The atom detection threshold of 5.015 photons (vertical dashed line) is determined by minimization of the sum loss and infidelities. We find $E_{av} = 0.031(3)$ and $E_{va} = 0.004(1)$.}
    \label{fig:ImagingHistogram}
\end{figure}

We determine $E_{av}$ and $E_{va}$ from measurement calibration data (see Fig.~\ref{fig:ImagingHistogram}), $E_{av,va} = m_{av,va}/\sum\textbf{m}$. The photon number threshold for atom-vacancy determination is chosen to minimize the sum of error processes. Having determined the measurement process matrix from the calibration data, on subsequent datasets, we load this matrix and apply its inverse to the vector of measured values to obtain the estimates of the true experimental events, $\textbf{N}$. Desired observables, such as loss between images or parity in a doublet, are then defined in terms of the elements of this vector.
\begin{align}
    \text{loss} &= \sum_i \frac{N_{av}}{N_{aa}+N_{av}} \nonumber \\
    \text{parity} &= \sum_i \frac{N^l_{aa}N^r_{aa} - N^l_{aa}N^r_{av} - N^l_{av}N^r_{aa} + N^l_{av}N^r_{av}}{N^l_{aa}N^r_{aa} + N^l_{aa}N^r_{av} + N^l_{av}N^r_{aa} + N^l_{av}N^r_{av}}
\end{align}
where the summation over $i$ is taken to refer to all sites or doublets in the array, respectively, and the index $i$ in the expressions is suppressed for clarity.

Determining the uncertainty in these measurements consists of two parts which are summed in quadrature, the statistical uncertainty and the SPAM correction uncertainty. The statistical uncertainty is determined by recognizing that the counts in the true event vector $\mathbf{N}$ are samples from a multinomial distribution. We assume that the probabilities $p_i = \mathbf{N}_i/\sum_i\mathbf{N}$ are good estimates of the probabilities in the multinomial distribution. The (co-)variances of the $\mathbf{N}_i$ are then determined as normal for a multinomial distribution and propagated through the expressions for the desired observables. 

The uncertainty associated with SPAM correction arises because the calibration of error probabilities are experimentally determined and therefore uncertain. Although it is relatively easy to take high statistics calibration data to make these uncertainties small, we nevertheless propagate the uncertainty as follows. First we take our measured uncertainties in $E_{va}$ and $E_{av}$ and convert those into an uncertainty in each of the 16 elements of $M^{-1}$. Given the measurement vector $\mathbf{m}$, which does not have associated uncertainty, we then find the uncertainties in the true event vector $\mathbf{N}$. Finally, these uncertainties in $\mathbf{N}$ are propagated through the expression for the desired observables. Finding the variances in the elements of $M^{-1}$ is, tricky. In principle, the expression for the measurement process matrix, Eqn. (\ref{eqn:measurementmatrix}), can be analytically inverted, and the uncertainty in the errors propagated directly. However, this expression is complex and quite difficult to work with. Instead, we treat the measurement process matrix as a matrix of random variables, where the mean and standard deviation of each entry is computed via Eqn. (\ref{eqn:measurementmatrix}) and standard propagation of uncertainty. We can take a large number (50,000) samples of the measurement process matrix distribution, invert each sample, and use the resulting distribution of each element of the inverse measurement process matrix to determine a mean and standard deviation for each element of the inverse process matrix. The mean value is unnecessary since we can just invert the measurement process matrix determined by our experimental measurements of the imaging loss and infidelities, but the standard deviations from this process provide an easy estimate of the uncertainties in this matrix. 

There are two significant notes of caution that must be made here. The first is that the inverse of a matrix of normal random variables is not itself a matrix of normal random variables. In general, the elements of the inverse of a matrix of normal random variables can have highly skew or bimodal distributions such that the standard deviation computed from sampling the inverse matrix is a poor estimate of the typical deviation from the the mean. However, for experimentally determined values of imaging loss and infidelity, this is not a major issue: The resulting distributions are somewhat skew but are not multi-modal, and so the computed standard deviation provides a decent estimate in the uncertainty in each element of the inverse measurement process matrix. The second note of caution is that each element of the measurement process matrix is not an independent random variable of known mean and standard deviation. In fact, we have two random variables which are combined in various ways to produce the measurement process matrix, resulting in significant covariances between its 16 elements.  

Typical total imaging error probabilities are between two and three percent. The measured uncertainty in these quantities is significantly smaller than the slow fluctuation in these quantities, which we add in as a systematic uncertainty throughout. Carefully addressing the concerns raised above will be necessary if we are to, in the future, quote tighter bounds with reduced systematic fluctuations. 

\section{Rydberg state loss yields systematic overestimate of $P_{gg}$}
\label{sec:RydSys}

The fidelity with which a Bell state is produced is given by $\mathcal{F} = \frac{1}{2}(P_{gg}+P_{ee}+C)$, where the populations $P_{gg}$ and $P_{ee}$ are measured directly after the application of the gate sequence, and $C$ is the contrast of the parity measured after a variable axis $\pi/2$-pulse following the gate sequence~\cite{sackett2000experimental}. We detect atoms in $\ket{e}$; our imaging does not distinguish between atoms which are in $\ket{g}$ and atoms which are lost. Our SPAM correction procedure accounts for loss arising due to imaging, as well as various trapping potential ramps and hold times. However, loss due to the operation of the gate itself, e.g. Rydberg state decay during the adiabatically resonant pulses or non-adiabaticity resulting in population remaining in the Rydberg state after the pulses, is not accounted for. Such loss causes decay of $\ket{ee}$ into $\ket{0e}$ or $\ket{e0}$, where $\ket{0}$ refers to a lost atom. These components, after the final $\pi/2$-pulse, can appear as $\ket{gg}$. The latter possibility is relevant since half of the state population before the Rydberg pulse is in these states and their decay causes an increase in measured $P_{gg}$, systematically inflating our measured fidelity. 

We correct for this by independently measuring the increase in $P_{gg} + P_{ee}$ caused by the coupling to the Rydberg state. More precisely, we perform the gate sequence without any blowaway pulse, so that our final image detects atoms in both $\ket{g}$ and $\ket{e}$. The absence of an atom therefore corresponds to true atom loss. By comparing populations after the gate sequence with and without the Rydberg coupling, we compute a systematic increase in $P_{gg}$ in the normal detection scheme of 1.8(8)\%. We correct for this by reducing our reported Bell state fidelity by 0.9(5)\%

In what follows, we outline the procedure to determine this systematic shift. We begin by defining the single-atom and two-atom Rydberg state loss or decay probabilities, $p_1$ and $p_2$. We extract these quantities from the calibration data of the gate sequence without blowaway detection as 
\begin{align}
p_1 &= 2\left( P(a|\text{no UV}) - P(a|\text{UV}) \right) \nonumber \\
p_2 &= 4\left( P(aa|\text{no UV}) - P(aa|\text{UV}) - \frac{p_1}{2} \right)
\end{align}
where in the first line, $P(a)$ refers to the (conditional) probability of measuring an atom present in the ensemble of single atoms, and in the second line $P(aa)$ refers to the (conditional) probability of measuring two atoms present in the ensemble of filled doublets. We find $p_1 = 0.011(6)$ and $p_2 = 0.075(24)$. We next determine how these loss probabilities affect the gate sequence with normal detection:
\begin{align}
P(gg) &= P(\text{no decay})P(gg | \text{no decay}) \,+ \nonumber \\
&\quad\,\, P(\text{decay})P(gg | \text{decay}) 
\end{align}
where $P(gg)$ is the probability of measuring both atoms in $\ket{g}$ in a fully filled doublet. We note that $P(\text{decay}) = p_1/2 + p_2/4$, and we label $P(gg | \text{no decay}) \equiv P_{gg}^0$. Finally, we determine that $P(gg | \text{decay}) = 0.75$ is the probability that the remaining atom after a Rydberg loss or decay event is measured to be in $\ket{g}$, after averaging over times within the Rydberg pulses that the decay may have occurred. This may be understood by taking limits. If the decay occurs at the beginning of the first pulse, the two atoms are in a product state, so the decay of one atom does not affect the other, which therefore proceeds like a single atom and is measured to be in $\ket{g}$. On the other hand, losing an atom out of a fully entangled state leaves the other atom in a mixed state of either $\ket{g}$ or $\ket{e}$. The final $\pi$/2-pulse rotates both of these possibilities into a an superposition, providing a 50\% chance of measuring that remaining atom to be in $\ket{g}$. Modeling the intermediate possibilities shows a smooth interpolation and provides the averaged expectation of $P(gg | \text{decay}) = 0.75$. Combining these observations, we have
\begin{align}
P_{gg} = (1-p_1/2-p_2/4)P_{gg}^0 + (p_1/2+p_2/4)\frac{3}{4}
\end{align}
We therefore measure a $\ket{gg}$ population that is increased by an amount $\frac{3}{16}(2p_1+p_2) = 0.018(5)$ due to Rydberg loss and decay. We also assign this a systematic uncertainty of $0.6\%$ by recognizing that $P(gg | \text{decay})$ is bounded between $0.5$ and $1$ based on the limits discussed above, and processes not considered here could result in a deviation from our best estimate of $0.75$. Therefore, we estimate the systematic increase of $P_{gg}$ to be 1.8(8)\%, so we reduce our measured Bell state fidelity by $0.9(4)\%$ to account for this effect.

Our measured Bell state fidelity is $\mathcal{F}_m= 93.7(2.0)\%$, so we report $\mathcal{F} = 92.8(2.0)\%$. An alternative lower bound on the fidelity may be found by using only the measured Bell state parity oscillation contrast, $C \le (P_{gg}+P_{ee})$. This provides a fidelity lower bound $\mathcal{F}_{lb} = C = 91.9(2.8)\%$. 

\section{Theory of Bell state fidelity}

In this section we write out explicitly the action of our gate sequence. In keeping with the main text, we label the electronic ground, clock, and Rydberg states $\ket{g}$, $\ket{e}$, and $\ket{r}$ respectively. Additionally the state $\ket{0}$ denotes the absence of an atom, or the population of a dark state such as $^3$P$_2$. For now we focus our attention on the ideal gate protocol, for which we can restrict our attention to the single-qubit basis $\{ \ket{e}, \ket{g} \}$, and instances of the experiment where we load both tweezers with an atom, initializing our system in the state $\ket{gg}$. With this notation, a clock rotation (i.e. transition between $\ket{g}$ and $\ket{e}$) will be denoted $r_\alpha (\phi)$ where the angle $\alpha$ parametrizes the axis of rotation on the Bloch sphere (which will be assumed to be in the $x-y$ plane) and $\phi$ is the rotation angle. A useful form for this operation is 

\begin{align}
r_\alpha (\phi) = \cos(\frac{\phi}{2}) \mathbbm{1} - i\sin(\frac{\phi}{2}) [\cos(\alpha) \sigma_x + \sin(\alpha) \sigma_y]
\end{align}
where $\sigma_x, \sigma_y$ are Pauli matrices and $\mathbbm{1}$ is the identity operator, all of which act on the basis $\{ \ket{e}, \ket{g} \}$. A global clock rotation on two qubits can be expressed as $R_\alpha (\phi) = r_\alpha (\phi) \otimes r_\alpha  (\phi) $. Next, we want to consider the effects of our adiabatic Rydberg-dressing scheme. In the absence of dissipation, the result of these dynamics can be understood by analyzing the lightshifts that they induce on the states in the $\ket{g}, \ket{e}$ basis, which lead to the following operation:

\begin{align}
\ket{gg} \ \ \ &\mapsto \ \ \  \ket{gg} \nonumber \\
\ket{eg \ \text{or} \ ge} \ \ \ &\mapsto \ \ \  e^{-i \theta} \ket{eg \ \text{or} \ ge} \nonumber \\
\ket{ee} \ \ \ &\mapsto \ \ \ e^{-i (2 \theta +  \Theta)} \ket{ee}.
\end{align}

Here $\theta$ is the accrued phase due to light shifts on the single-atom state $\ket{e}$. In the non-interacting limit, the two-atom state $\ket{ee}$ picks up twice the single-particle shift, $2\theta$. However, in the blockaded regime, the light shift on $\ket{ee}$ is different from the light shift due to single-particle effects by the entangling energy $\kappa$, as discussed in the main text. As such, the final phase accrued by $\ket{ee}$ will differ from single-particle expectation $2\theta$ by an amount $\Theta$.

\begin{align}
U(\theta, \Theta) = e^{-i\frac{\theta}{2}} &\bigl{[} \cos(\theta/2) \mathbbm{1} \otimes \mathbbm{1} + i \sin(\theta/2) \sigma_z \otimes \sigma_z \bigr{]} + \nonumber\\ 
&\bigl{[} e^{-i(2\theta + \Theta)} - 1\bigr{]} \ket{ee}\bra{ee}.
\end{align}
In terms of these two experimentally realizable operations $R$ and $U$, the full entangling gate operation $G$ is

\begin{align}
G = R_x (\pi/2) U(\theta_2,  \Theta_2) R_x (\pi) U(\theta_1,  \Theta_1) R_x(\pi/2) \label{eq:full_gate},
\end{align}
which is equivalent to the schematic sequence in figure~\ref{fig:gate}d, with the addition that the accrued phases from adiabatic dressing are subscripted in each pulse so as to capture the effect of differential laser-amplitude noise between the two ramps.

As a metric for fidelity, we can calculate the overlap of the generated state with that of  the target Bell state $ \ket{\psi_B} = \frac{1}{\sqrt{2}} \bigl{[} \ket{gg} + i\ket{ee} \bigr{]}$. Expressing the density matrix for the two-atom pair after the gate sequence by $\rho$, the fidelity is $\mathcal{F} = \bra{\psi_B}\rho\ket{\psi_B}$. If we define $\delta\theta = \theta_1 - \theta_2$, $\delta\Theta = \Theta_1 - \Theta_2$ and $\Theta = (\Theta_1 + \Theta_2)/2$ this can be written

\begin{align}
\mathcal{F} = \frac{1}{4}\biggl{(} 1+&\cos(\delta\theta + \frac{1}{2}\delta\Theta)^2 + 2\cos(\delta\theta + \frac{1}{2}\delta\Theta)\sin(\Theta) \biggr{)}
\end{align}
From this expression, we see that we want to tune the parameters of our adiabatic ramps such that $\Theta = \pi/2$ and $\delta\theta = \delta\Theta = 0$ in order to prepare the desired Bell state.

\section{Numerical estimation of Bell state fidelity}
\label{SI:mastereqn}

In practice, the fidelity with which we can prepare and measure a Bell state is limited by numerous experimental considerations. In this section, we numerically characterize the expected infidelity due to Rydberg state decay, non-adiabaticity in our dressing protocol, laser noise, and single-qubit rotation errors. In order to efficiently simulate these sources of error, we treat the operations of $U$ and $R$ separately. In particular, we act on our state with $R$ according to unitary Schr\"{o}dinger evolution. However, we treat the presence of imperfect clock rotations by developing a specific noise model for the error in rotation angle. In order to simulate each adiabatic ramp $U$, we use a master-equation-based model where we consider two 4-level-systems, each spanned by $\{ \ket{e}, \ket{g}, \ket{r}, \ket{0} \}$, and the evolution of a density matrix $\rho$ according to 

\begin{align}
\dot{\rho} = -i[H(t),\rho] + \gamma_r \mathcal{L}(\rho, J_r)
\end{align}

\begin{align}
H(t) = -\Delta(t) &\Bigl{[} \ket{r} \bra{r} \otimes \mathbbm{1} +  \mathbbm{1} \otimes \ket{r} \bra{r} \Bigr{]} + \nonumber\\
\frac{\Omega_r(t)}{2} &\Bigl{[} \ket{r} \bra{e} \otimes \mathbbm{1} +  \mathbbm{1} \otimes \ket{r} \bra{e} + \text{h.c.} \Bigr{]} + \nonumber \\
V_{DD} &\ket{rr} \bra{rr}
\end{align}

\begin{align}
\mathcal{L}(\rho, J_r) = J \rho J_r^\dagger - \frac{1}{2} ( \rho J_r^\dagger J_r + J_r^\dagger J_r \rho)
\end{align}
where $H(t)$ is our time-dependent Hamiltonian, which includes adiabatic ramps of detuning $\Delta(t)$ and Rabi frequency $\Omega_r(t)$. $J_r = \ket{0}\bra{r}$ is the jump operator associated with decay of the Rydberg state to $\ket{0}$, and has a rate $\gamma_r$. The following sections discuss how we incorporate noise into this model, and develop a partial error budget for our experiment. 

\begin{figure}[t]
	\centering
	\includegraphics[width=0.9\columnwidth]{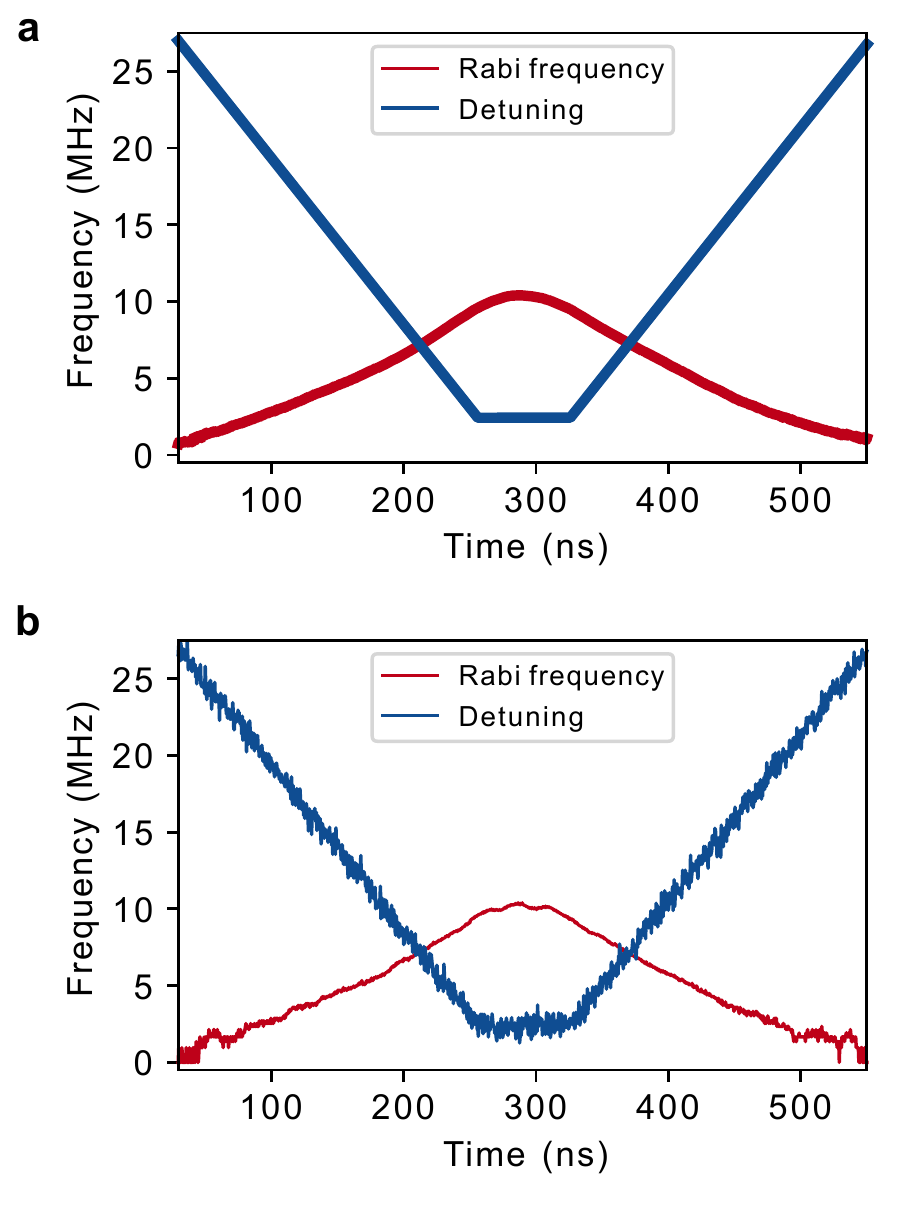}
	\caption{\textbf{Rydberg pulse ramps. a)} In red (blue) we plot the average (idealized) ramp sequence for the Rabi frequency (detuning) of our 317 nm Rydberg transition laser. This protocol adiabatically prepares a strongly-dressed eigenstate, which is a superposition of the states $\ket{e}$ and $\ket{r}$. \textbf{b)} Plotted in red (blue) are measured (simulated) single-shot instances of our Rabi-frequency (detuning) ramps.}
	\label{fig:ramps}
\end{figure}

\begin{figure*}
	\centering
	\includegraphics[width=0.8\textwidth]{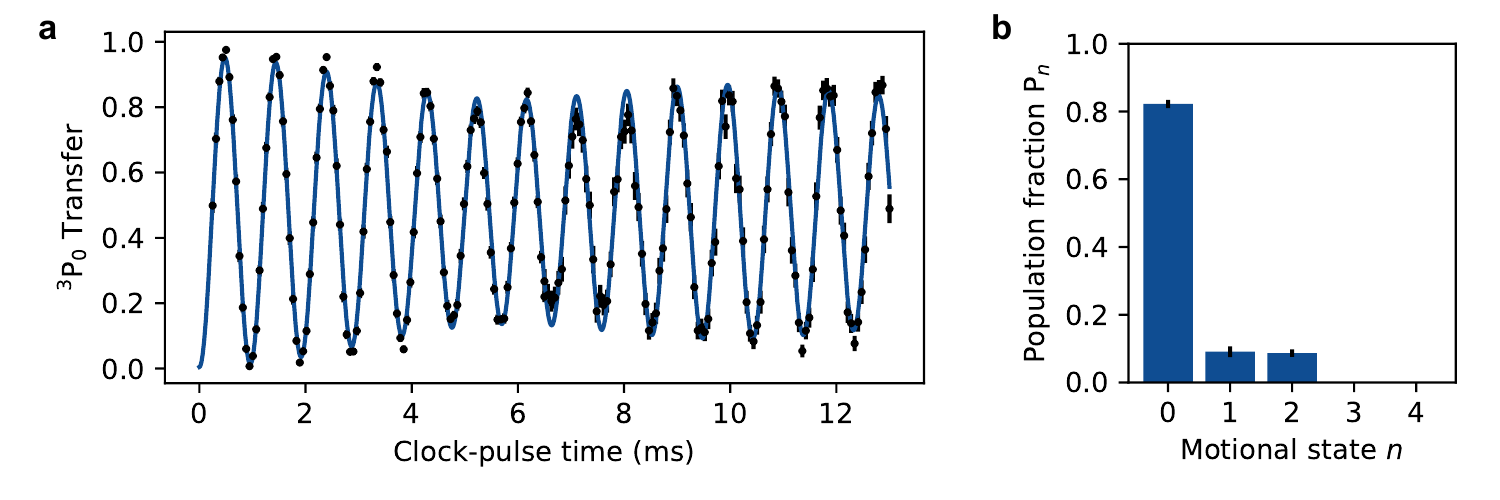}
	\caption{\textbf{Clock Rabi oscillations. a)} Rabi oscillation amplitude beats over tens of cycles due to increased atomic temperature. With the functional form in Eq. \ref{eq:rabi_motional}, we fit $\Omega_c = 2\pi \times 1.055(1)$ kHz. \textbf{b)} The beating pattern in our Rabi oscillations is set by the different frequencies of each motional state. In the fit to Eq. \ref{eq:rabi_motional}, we use $n_\text{max} = 4$ and infer the motional state population fractions. Error bars represent 95\% confidence intervals.}
	\label{fig:clock_rabi_beating}
\end{figure*}

\subsubsection{Non-adiabaticity}

In order to isolate the magnitude of errors due to non-adiabaticity, we numerically simulate our experiments using measured laser intensity traces, from which we can deduce the form of our Rabi frequency ramp. For the linear ramps of the detuning $\Delta(t)$, we assume an idealized profile. These Rabi frequency and detuning ramps are shown in figure~\ref{fig:ramps}a. We then optimize the simulated Bell state fidelity of the final state over a realistic range of final Rabi frequencies in our intensity ramp, and of frequency offsets in our ramp of detuning, as would be done in a typical experiment. Additionally, we assume perfect clock rotations between Rydberg dressing ramps, and infinite state lifetimes. Based on this procedure, we expect that non-adiabaticity causes 0.97\% infidelity in our Bell state preparation. 

\subsubsection{Rydberg state lifetime}

To include the effects of the finite Rydberg state lifetime, we set the parameter $\gamma_r$ (which was assumed to be 0 in the previous section) to correspond to the decay rate of the state $\ket{r}$, and compare the new Bell state fidelity to that achieved with infinite state lifetimes. This allows us to infer a lifetime related infidelity of 0.94\%. Propagating the uncertainty on our estimate of the Rydberg state lifetime, we also calculate a 95\% confidence interval of [0.83\%, 1.07\%] for this error. 

\subsubsection{Intensity noise}

Next, we repeat our master equation simulations, including a realistic Rydberg state lifetime, but with measured traces for the Rabi frequency $\Omega_r(t)$ as shown in Fig.~\ref{fig:ramps}b. For this simulation, we use the same gate parameters that were found to be optimal for the average Rabi frequency ramp, and do not perform any additional optimization. By averaging the resulting fidelity over many such simulations with different individual laser intensity traces before and after the spin echo, we find an average infidelity of 0.73\% due to this effect. Additionally, we can perform a bootstrapping analysis on the resulting distribution of infidelities, and calculate a 95\% confidence interval of [0.56\%, 0.90\%] on this statistic.

\subsubsection{Phase noise}

In order to capture the effect of laser phase noise, we return to using the mean Rabi frequency ramp, but now add frequency noise to our detuning ramps. A representative example is shown in figure~\ref{fig:ramps}b. We calculate many representative laser noise traces by measuring the frequency noise spectrum of our 317 nm laser. This is done by analyzing the in-loop PDH error signal of the undoubled 634 nm light locked to a ULE cavity. The inferred noise spectrum for our 317 nm laser is then used to generate frequency noise traces, which we add to the idealized detuning ramp in figure~\ref{fig:ramps}a. As was done with the measured Rabi frequency traces, we use our simulation to calculate the expected level of infidelity due to frequency noise, and find that, for our gate parameters, it introduces a negligible ($<$0.03\%) level of infidelity. Again performing a bootstrapping analysis, we find a 95\% confidence interval of [-0.04\%, 0.03\%]. The negative bound in this confidence interval corresponds to the fact that some noise traces actually improve the final Bell state fidelity. This is not in tension with the fact that we optimize the final detuning of the frequency ramp, because the noise traces have frequency components that are significantly higher than what we can control. As such, it is possible that high frequency alterations to the adiabatic frequency sweep may deform the ramp into a slightly more optimal form. 

\subsubsection{Imperfect clock-rotations}

Our gate protocol relies on the ability to perform high fidelity single-qubit rotations. In order to estimate our ability to perform these rotations, we can look at the decay of our Rabi oscillations on this transition, which effectively amounts to the repeated application of single-qubit operations with coherent errors. The results are shown in figure~\ref{fig:clock_rabi_beating}a. From this measurement we observe a beating pattern with high contrast revivals even after many oscillations. This allows us to identify the dominant mechanism for rotation errors as differential effective Rabi frequencies for atoms in different motional states of our trap. More precisely, we model the Hamiltonian during our clock-Rabi oscillations as

\begin{align}
H_{\text{Clock-Rabi}} = \frac{\Omega_c}{2} \bigl{[} e^{i\eta(a^\dagger + a)} \frac{\sigma_x + i \sigma_y}{2} + \text{h.c.}\Bigr{]}.
\end{align}
In this equation $\Omega_c$ is the Rabi frequency, $\eta$ is the Lamb-Dicke parameter, $a$ ($a^\dagger$) is the motional lowering (raising) operator, and h.c. denotes the Hermitian conjugate of the preceding term. The effective Rabi frequency for an atom in a given motional state $n$ is then 

\begin{align}
\tilde{\Omega}_c(n) = \Omega_C e^{-\eta^2 / 2} L_n(\eta^2),
\end{align}
where $L_n$ is the Laguerre polynomial of degree $n$. The expected functional form for Rabi oscillations with atoms in a distribution of motional states is

\begin{align}
P_{\ket{e}}(t) &= A \sum_{n=0}^{n_\text{max}} P_n \sin^2\Biggl{(}\frac{\tilde{\Omega}_C(n) t}{2} \Biggr{)}, \label{eq:rabi_motional}
\end{align}

\begin{align}
1 = \sum_{n=0}^{n_\text{max}} P_n.
\end{align}

By fitting our observed Rabi oscillations with Eq. \ref{eq:rabi_motional} and a fixed $\eta = 0.22$, set by independent measurements of trap frequencies, we find that for our ensemble-averaged oscillations, $\Omega_C = 2\pi \times 1.055(1)$ kHz. Finally, the inferred distribution of motional states is shown in figure Fig. \ref{fig:clock_rabi_beating}b. We note that the data in figure~\ref{fig:clock_rabi_beating}a are not SPAM corrected, as this does not affect the fitted Rabi frequency or inferred distribution of motional states.

We can now model errors in our clock rotations by sampling the effective Rabi frequency from a discrete set corresponding to the Rabi frequencies of each motional state, and with a probability corresponding to the population fractions plotted in figure~\ref{fig:clock_rabi_beating}b. By fixing the clock pulse times that correspond to $\pi$ and $\pi/2$ rotations, this gives a random rotation angle for each single-qubit rotation. Including this in our simulation introduces an added infidelity of 0.57\%, with a 95\% confidence interval of [0.39\%, 0.77\%].

\subsubsection{Error budget}

Here we tabulate the contributions to Bell state infidelity. To obtain a total expected infidelity, we repeat our simulation including all listed sources of error, and calculate a total expected infidelity of 3.20\%, with a 95\% confidence interval of [2.84\%, 3.60\%]. The confidence interval of the total error is obtained by a bootstrapping analysis to take into account correlations between error sources. To this, we add in our measured systematic overestimate of $P_{gg}$ due to Rydberg state loss (see~\ref{sec:RydSys}). Our reported fidelity is reduced by this amount, so we add this into our error budget.

\begin{center}
	\begin{tabular}{ |c|c|c|c| } 
		\hline
		Effect & Error & 95\% Confidence interval \\
		\hline
		Non-adiabaticity &0.97\% & $-$\\ 
		Rydberg state lifetime & 0.94\% & [0.83\%, 1.07\%] \\ 
		Intensity noise & 0.73\% & [0.56\%, 0.90\%] \\ 
		Phase noise & 0.00\% & [-0.04\%, 0.03\%] \\ 
		Clock rotation errors & 0.57\% & [0.39\%, 0.77\%] \\ 
		Rydberg decay systematic & 0.90\% & $-$\\
		\hline
		\textbf{Total:} & \textbf{4.10\%} & [3.74\%, 4.50\%] \\ 
		\hline
	\end{tabular}
\end{center}

\subsubsection{Interdoublet interactions}

Stray interactions between two separate doublets should be negligible based on simulations of interaction energies at the minimum, six lattice site separation between two doublets~\cite{robertson2021arc}. However, to investigate if undesired interactions between doublets is nevertheless limiting the Bell state fidelity, we perform Bell state fidelity experiments in a 4$\times$5 doublet array, where the spacing between doublets is increased from six lattice sites to eleven. Comparing datasets taken back-to-back we find that larger doublet spacing increases Bell state fidelity by 1.6(2.6)\%. Because this is consistent with zero and Rydberg simulations predict negligible interactions at normal separations, we do not add in expected infidelity from this source. 

\subsubsection{Lattice loading slips}

As discussed in supplement section~\ref{SI:apparatus}, we observe lattice loading `slips', wherein an atom is loaded into a lattice site adjacent to its target; in our Bell state fidelity dataset, we find $\sim$10\% of atoms experience a slip. On average, one quarter of these atoms will be displaced directly away from its partner within the doublet, and our numerical model indicates that a separation increase from two lattice sites to three lattice sites results in a breakdown of blockade and poor Bell state fidelity for 5s40d $^3$D$_1$ Rydberg states. This would indicate a reduction in Bell state fidelity due to slips to be at the percent level. We have attempted advanced analysis techniques to post-select for atoms that have not slipped, but this has not significantly affected the measured Bell state fidelity. But, this analysis is complicated by removing the normally-used assumptions on where the atoms will be based on intended loading locations, and the combination of our imaging point-spread function and atom detection signal makes slip detection in a single shot imperfect. As a result, this remains a plausible additional source of error that could contribute to the difference between our error budget and measured infidelity.

\newpage
\onecolumngrid

\section{Connection of correlations to Bell state coherence}
We explore the Bell state coherence lifetime via two distinct correlation measurements: the parity-parity correlation within the ensemble of Bell states, and the parity-$S_z^2$ correlation between the Bell state ensemble and single atom ensemble. Here we show how the Bell state coherence is connected to these correlators. 

\subsection{parity-parity correlation}

We consider an arbitrary density matrix for two two-level-atoms. 

\begin{equation}
\rho_2=\left(
\begin{tabular}{cccc}
$p_{gg}$ & $c_{gg,ge}e^{i\phi_{gg,ge}}$ & $c_{gg,eg}e^{i\phi_{gg,eg}}$ & $c_{gg,ee}e^{i\phi_{gg,ee}}$ \\  
$c_{gg,ge}e^{-i\phi_{gg,ge}}$ & $p_{ge}$ & $c_{ge,eg}e^{i\phi_{ge,eg}}$ & $c_{ge,ee}e^{i\phi_{ge,ee}}$  \\ 
$c_{gg,eg}e^{-i\phi_{gg,eg}}$ & $c_{ge,eg}e^{-i\phi_{ge,eg}}$ & $p_{eg}$ & $c_{eg,ee}e^{i\phi_{eg,ee}}$ \\ 
$c_{gg,ee}e^{-i\phi_{gg,ee}}$ & $c_{ge,ee}e^{-i\phi_{ge,ee}}$ & $c_{eg,ee}e^{-i\phi_{eg,ee}}$ & $p_{ee}$ \\ 
\end{tabular} 
\right)
\label{Eqn:2atomdensitymatrix}
\end{equation}

An ideal Bell state has $p_{gg} = p_{ee} = c_{gg,ee} = 1/2$, with arbitrary but fixed coherence phase $\phi_{gg,ee}$, and all other entries zero.

To characterize Bell state lifetimes, after preparing the Bell states and waiting through the hold time, we perform a final analysis $\pi/2$-pulse about a variable axis. The analysis pulse implements the operator $R_{\theta}(\pi/2)$ given by
\begin{align}
R_{\theta}(\pi/2) &= R_{x}(\pi/2)R_z(\theta) \nonumber \\
R_{\alpha}(\theta) &= \exp(-i \hat{S}_\alpha\,\theta) 
\end{align}
where $\hat{S}_\alpha = \frac{1}{2}\left(\hat{\sigma}_\alpha\otimes\mathbb{1}\otimes\mathbb{1}\otimes... \,+\, \mathbb{1}\otimes\hat{\sigma}_\alpha\otimes\mathbb{1}\otimes... \,+\, ... \right)$ and $\hat{\sigma}_\alpha$ are the Pauli matrices. In the context of two atoms in a doublet, this reduces to $\hat{S}_\alpha = \frac{1}{2}(\hat{\sigma}_\alpha\otimes\mathbb{1} + \mathbb{1}\otimes\hat{\sigma}_\alpha)$. Note that the implemented analysis pulse would ideally have the form $R_{\theta}(\pi/2) = R_z(-\theta)R_{x}(\pi/2)R_z(\theta)$; however, the second $z$-rotation does not affect any of our results and we do not include it to reduce the number of physically implemented pulses.	

Finally, we measure the parity of each Bell state, $\hat{\Pi} = \hat{\sigma}_z \otimes \hat{\sigma}_z$.
\begin{align}
\Pi(\theta) &= \Tr\left(R_{\theta}(\pi/2) \rho_2 R_{\theta}^\dagger(\pi/2) \hat{\Pi}\right) \nonumber \\
&= 2\left(c_{gg,ee}\cos(2\theta-\phi_{gg,ee}) + c_{ge,eg}\cos(\phi_{ge,eg}) \right)
\label{Eqn:paritysig}
\end{align}
We note that no populations and only two coherences appear. The first, $c_{gg,ee}$ is the only coherence that appears for our Bell state. Both coherences could arise if the supposed Bell state were in fact a product state of single atoms in a superposition, in which case $c_{ge,eg} = c_{gg,ee} = c_{g,e}^2$, $\phi_{ge,eg} = 0$, and $\phi_{gg,ee} = 2\phi_{g,e}$, where $c_{g,e}$ and $\phi_{g,e}$ are the magnitude and phase of the single atom density matrix coherence. 

Our expected parity-parity correlation is obtained from this parity signal from two Bell states by appropriate averaging over final analysis pulse angle, $\theta$. 

\begin{align}
\mathcal{C}_{\Pi\Pi} &= \left\langle\left(\hat{\Pi}-\bar{\Pi}\right)_i\left(\hat{\Pi}-\bar{\Pi}\right)_j\right\rangle \nonumber\\
&= \left\langle \hat{\Pi}_i \hat{\Pi}_j \right\rangle - \langle\hat{\Pi}_i\rangle\langle\hat{\Pi}_j\rangle \nonumber \\
&= \frac{1}{2\pi}\int_0^{2\pi} \Pi^i(\theta)\,\, \Pi^j(\theta)  \diff{\theta} - \left(\frac{1}{2\pi} \int_0^{2\pi} \Pi^i(\theta) \diff{\theta} \right) \left(\frac{1}{2\pi} \int_0^{2\pi} \Pi^j(\theta) \diff{\theta}\right)
\label{Eqn:correlatorform}
\end{align}
where we do not assume that the average value of both signals $i$,$j$ is identical, since the coherence magnitudes and phases may differ between Bell states. Nevertheless, plugging Eqn. (\ref{Eqn:paritysig}) into Eqn. (\ref{Eqn:correlatorform}), we find

\begin{align}
\mathcal{C}_{\Pi\Pi}&= 2 c_{gg,ee}^2\cos(\phi_{gg,ee}^i-\phi_{gg,ee}^j) 
\end{align}

This indicates that the parity-parity correlation decays at twice the rate of the Bell state coherence decay. This correlation function is also sensitive to product of superposition states with reduced amplitude: the maximum parity-parity correlation for a Bell state is $1/2$ while the maximum correlation for a product state is $1/8$. Because we separately determine our initial Bell state fidelity, and decay from a Bell state directly to a product of superposition states is improbable, we use this correlation function as a measure of the Bell state coherence time. 	

\subsection{parity-$S_z^2$ correlation}

The parity signal that goes into the parity-$S_z^2$ correlation function has been computed in Eqns. (\ref{Eqn:2atomdensitymatrix}) and (\ref{Eqn:paritysig}). For the $S_z$ signal, we follow a similar structure as before, starting by considering a general single atom density matrix,

\begin{equation}
\rho_1=\left(
\begin{tabular}{cc}
$p_g$ & $c_{g,e}e^{i\phi_{g,e}}$ \\  
$c_{g,e}e^{-i\phi_{g,e}}$ & $p_e$
\end{tabular} 
\right).
\end{equation}

The analysis pulse applies a $\pi/2$-pulse about a variable axis, implementing the same operator as in the previous sections except that here $\hat{S}_\alpha = \frac{1}{2}\hat{\sigma}_\alpha$ since we are considering a single atom Hilbert space. We measure the single-experiment array-averaged spin projection, $S_z$, given by

\begin{align}
S_z(\theta) &= \Tr\left((R_{\theta}(\pi/2) \rho_1 (R_{\theta}^\dagger(\pi/2) \hat{S}_z \right) \nonumber \\
&= -c_{g,e}\cos(\theta-\phi_{g,e})
\label{Eqn:Szsignal}
\end{align}

Our expected parity-$S_z^2$ correlation is obtained by combining the single-experiment array-averaged spin projection signal, $S_z$ with the single-experiment array-averaged Bell state parity signal $\Pi$, with appropriate averaging over analysis pulse angle, $\theta$.

\begin{align}
\mathcal{C}_{\Pi S_z^2} &= 4 \langle (\Pi-\bar{\Pi}) (S_z^2-\overline{S_z^2}) \rangle \nonumber \\
&= 2 c_{g,e}^2 c_{gg,ee}\cos(2\phi_{g,e}- \phi_{gg,ee})
\end{align}

where to obtain the second line, we have used Eqns. (\ref{Eqn:paritysig}) and (\ref{Eqn:Szsignal}), and we have assumed that the density matrices are uniform within each ensemble.

This indicates that the parity-$S_z^2$ correlation decays at the sum of the decay rate of the Bell state coherence and twice the decay rate of the single-atom coherence. The same data set provides both the decay of this correlation function and the decay of the $\hat{S}_z$-$\hat{S}_z$ correlation that directly provides twice the decay rate of the single-atom coherence. 
\end{document}